\documentclass[aps,prc,reprint,superscriptaddress]{revtex4-1}
\usepackage{graphicx}
\graphicspath{{./graphics/}}  %Folder containing figure (latex will search for this)
\usepackage{amsmath}
\usepackage{array}
\newcolumntype{+}{>{\global\let\currentrowstyle\relax}}
\newcolumntype{^}{>{\currentrowstyle}}
\newcommand{\rowstyle}[1]{\gdef\currentrowstyle{#1}%
   #1\ignorespaces
}
\def\bfseries{\fontseries \bfdefault \selectfont \boldmath}

\usepackage{scrextend}
\usepackage{hyperref}
\newcommand{\comment}[1]{}
\newcommand{\degree}{\ensuremath{^\circ}}
\usepackage{rotating}          % for sideways tables
\newcommand{\gtaeq}{\lower 2pt \hbox{$\, \buildrel {\scriptstyle >}\over {\scriptstyle \sim}\,$}}
\newcommand{\ltaeq}{\lower 2pt \hbox{$\, \buildrel {\scriptstyle <}\over {\scriptstyle \sim}\,$}}

\usepackage[export]{adjustbox} 

\begin{document}

%\preprint{}

%TODO
%1. Introduction:  DONE!!
%2. Experiment:  DONE!!
%3. Analysis
%	a. Intro to obtaining flux-averaged cross-sections:  DONE!!
%       b. Neutron attenuation:  DONE!!
%	c. GEANT4 simulations for determining efficiencies:
%       d. Final flux-averaged cross-sections
%	e. Background estimate:
%4. Conclusion:
%5. Acknowledgements:
%6. References:

\title{Cosmogenic-neutron activation of TeO\textsubscript{2} and implications for \\neutrinoless double-beta decay experiments}


\author{B. S. Wang}
%\thanks{}
\affiliation{Department of Nuclear Engineering, University of California, Berkeley, California 94720, USA}

\author{E. B. Norman}
\affiliation{Department of Nuclear Engineering, University of California, Berkeley, California 94720, USA}
\affiliation{Lawrence Livermore National Laboratory, Livermore, California 94550, USA}

\author{N. D. Scielzo}
\affiliation{Lawrence Livermore National Laboratory, Livermore, California 94550, USA}

\author{A. R. Smith}
\affiliation{Lawrence Berkeley National Laboratory, Berkeley, California 94720, USA}

\author{K. J. Thomas}
\affiliation{Department of Nuclear Engineering, University of California, Berkeley, California 94720, USA}
\affiliation{Lawrence Berkeley National Laboratory, Berkeley, California 94720, USA}

\author{S. A. Wender}
\affiliation{Los Alamos National Laboratory, Los Alamos, New Mexico 87545, USA}

% Got these affiliations from the CUORE sensitivity paper
%\author{M. Pavan}
%\affiliation{Dipartimento di Fisica, Universit\`{a} di Milano-Bicocca, Milano I-20126, Italy}
%\affiliation{INFN - Sezione di Milano Bicocca, Milano I-20126, Italy}

%\author{S. Capelli}
%\affiliation{Dipartimento di Fisica, Universit\`{a} di Milano-Bicocca, Milano I-20126, Italy}
%\affiliation{INFN - Sezione di Milano Bicocca, Milano I-20126, Italy}


\date{\today}

\begin{abstract}
%CUORE is an experiment that will search for neutrinoless double-beta
%($0\nu\beta\beta$) decay of \textsuperscript{130}Te  
%using an array of 988 high-resolution TeO\textsubscript{2} bolometers.
%Observation of the $0\nu\beta\beta$ decay signature requires ultra-low 
%background rates.  Therefore analysis of all possible background sources
%is essential.
%Background-source identification and characterization are therefore
%extremely important.

%This work examines one source of background that is poorly characterized in
%CUORE: cosmogenic neutron activation of TeO\textsubscript{2}.
%A measurement has been performed in which a sample of TeO\textsubscript{2} powder
%was irradiated with a neutron spectrum similar to that of cosmic-ray neutrons
%at sea-level.   
Flux-averaged cross sections for cosmogenic-neutron activation of 
natural tellurium were measured using a neutron beam containing neutrons of 
kinetic energies up to $\sim$800 MeV, and having an energy spectrum 
similar to that of cosmic-ray neutrons at sea-level.
Analysis of the radioisotopes produced reveals that \textsuperscript{110m}Ag 
will be a dominant contributor to the cosmogenic-activation background in 
experiments searching for neutrinoless double-beta decay of 
\textsuperscript{130}Te, such as CUORE and SNO$+$. 
An estimate of the cosmogenic-activation background in the CUORE 
experiment has been obtained using the results of this measurement and 
cross-section measurements of proton activation of tellurium. 
Additionally, the measured cross sections in this work are also compared with 
results from semi-empirical cross-section calculations.

\end{abstract}

%\keywords{}

\maketitle{}

\section{Introduction}
%The search for neutrinoless double-beta ($0\nu\beta\beta$) decay plays an  
%integral part in answering fundamental questions about particle physics
%and cosmology.
%The search for neutrinoless double-beta ($0\nu\beta\beta$) plays an  
%integral part in answering fundamental questions about the nature 
%of neutrinos.  
Neutrinoless double-beta ($0\nu\beta\beta$) decay \cite{Avignone2008,Rodejohann2011,Bilenky2012} 
is a long sought-after second-order
weak process in which a nucleus (A,Z) transitions to a nucleus (A,Z+2) through the
emission of two electrons.  This process is hypothesized to occur only if 
neutrinos are Majorana particles.  Observation of $0\nu\beta\beta$ decay would not 
only establish that neutrinos are Majorana fermions, but may also constrain the 
neutrino-mass scale and hierarchy, and demonstrate that total lepton number is not 
conserved. 
%Lepton-number violation may help explain the asymmetry between matter and 
%antimatter in the universe.

In experiments searching for $0\nu\beta\beta$ decay, the signature of interest is a
peak at the double-beta decay Q value (Q\textsubscript{$\beta\beta$}). 
As $0\nu\beta\beta$ decay would be a rare process,
%In $0\nu\beta\beta$-decay experiments, 
%full characterization of the background is
minimizing the background rate around %the $\beta\beta$-decay Q value 
%(Q\textsubscript{$\beta\beta$}) is essential
Q\textsubscript{$\beta\beta$} is essential
for improving the experimental sensitivity.
Therefore, a detailed characterization of all potential sources of background is 
important, as any event that can mimic or obscure the $0\nu\beta\beta$-decay peak 
%signature, a peak at Q\textsubscript{$\beta\beta$}, 
is problematic and must be 
well-understood and, if possible, eliminated.  
%Minimizing the background rate is essential for improving the
%experimental sensitivity, i.e., the lower limit one can set on the 
%$0\nu\beta\beta$-decay partial half-life when a null result is observed \cite{2011_Alessandria?}.  
%For example, in experiments where the source is part of the active region of the 
%detector, the sensitivity ($\widehat{T^{0\nu\beta\beta}_{1/2}}$) is inversely 
%proportional to the square root of the background rate ($b$) in the region 
%where the $0\nu\beta\beta$-decay signature is expected:
%of a $0\nu\beta\beta$-decay experiment 
%can be taken to be 
%is defined as 
%the lower limit one can set on the $0\nu\beta\beta$-decay partial half-life 
%when a null result is observed \cite{2011_Alessandria?}.  For experiments in which
%the source is part of the active region of the detector, this lower limit is
%
%\begin{equation}
%  \widehat{T^{0\nu\beta\beta}_{1/2}} \propto a\sqrt{\frac{M\cdot t}{b\cdot \delta E}}.
%  \label{eq:ExperimentalSensitivity}
%\end{equation}
%Here, $a$ is the isotopic abundance of the source isotope, $M$ is the total active mass,
%$t$ is the total live time of the experiment, and $\delta E$ is the energy resolution
%at the neutrinoless double-beta decay Q-value.  The rate $b$ is in counts/(keV$\cdot$kg$\cdot$y).

To miminize external backgrounds, $0\nu\beta\beta$-decay experiments operate 
in underground laboratories, where large overburdens decrease the flux of 
cosmic rays by orders of magnitude relative to the flux above ground \cite{Aglietta1998}.  
Further reduction of the remaining cosmic-ray background can be achieved with muon-veto detectors,  
and backgrounds from natural radioactivity in the laboratory environment can be
alleviated with proper shielding.
%Background also comes from neutrons and gamma rays from natural radioactivity in the 
%laboratory environment, which can be alleviated with proper shielding. 
%can alleviate this issue.     

Radioactivity present within the detector itself can provide a source of
background that is difficult to eliminate.  $0\nu\beta\beta$-decay 
experiments devote a great deal of effort into making ultraclean and ultrapure 
detector materials free of primordial radioisotopes.
However, no matter how clean or purely produced the materials are, 
cosmogenic activation will generate some radioactivity while the materials 
are at or above the Earth's surface during storage, production, or 
transportation \cite{Cebrian2006,Elliott2010,Lozza2014}.  The background contribution from this radioactivity can be 
minimized by ensuring detector materials spend as little time above ground as 
possible and by avoiding air transportation, as the
cosmic-ray flux increases significantly at higher altitudes \cite{Hess1959,Goldhagen2004}.  
%is higher in the atmosphere, e.g., the cosmic-ray-neutron flux at an 
%altitude of 12 km is approximately two orders of magnitude higher than the flux at 
%sea-level \cite{1959_Hess}.  
At sea-level, activation is primarily caused by the hadronic component of 
the cosmic-ray flux, which is dominated by neutrons \cite{Heusser1995}.

This work investigates the backgrounds associated with cosmogenic activation of
tellurium, which are important to understand for  
%a subject where data has been lacking but is important for 
%that, 
%to date, 
%due to a lack of data, have been poorly characterized 
%in 
experiments such as 
the Cryogenic Underground Observatory for Rare Events (CUORE) \cite{CUORE2004} and the 
Sudbury Neutrino Observatory Plus (SNO$+$) \cite{Hartnell2012} that are searching 
for the $0\nu\beta\beta$ decay of \textsuperscript{130}Te,
but to date are poorly characterized due to a lack of data.
As $0\nu\beta\beta$-decay experiments run for several years, 
%and Q\textsubscript{$\beta\beta$} is 2528 keV for \textsuperscript{130}Te,
typically only long-lived cosmogenic isotopes (i.e., that have half-lives 
of order a year or longer) with Q values near or greater than the
\textsuperscript{130}Te Q\textsubscript{$\beta\beta$} of 2528 keV \cite{Redshaw2009,Scielzo2009,Rahaman2011,Nesterenko2012} 
will be potential sources of background at the $0\nu\beta\beta$-decay peak.     
%and produce (or activate) isotopes. 

Determining the resulting cosmogenic-activation background contribution to a 
$0\nu\beta\beta$-decay experiment %can therefore be estimated from  
%by examining 
requires estimating the production rates of the radioisotopes in tellurium. 
%that can be produced by cosmogenic-neutron interactions with tellurium.
%of this activation to the background of a $0\nu\beta\beta$-decay
%experiment can be estimated by examining the production rates of all radioisotopes that 
%can be produced in tellurium by cosmogenic neutrons.
%The impact activation of TeO\textsubscript{2} will have on CUORE can be 
%understood by examining all cosmogenically-produced radioisotopes and 
%calculating their production rates.
Activation cross sections that span a wide range of neutron 
energies, from thermal up to several GeV, are therefore needed; however,
experimentally-measured cross-section data is currently sparse. 
%for the interactions of interest.
%At high neutron energies ($\gtaeq 100$ MeV) cross sections for neutron 
For neutron energies above 800 MeV, cross sections for neutron 
activation are expected to be approximately equal to those for proton 
activation, and can be estimated from existing experimental data for proton 
energies 800 MeV -- 23 GeV 
\cite{Bardayan1997,Norman2005,Barghouty2013}.
%estimated from measured proton-activation data for natural tellurium 
%\cite{Barghouty2013-ProtonsOnTe}.  
%Experimental data for proton-activation of tellurium exists for proton
%energies 800 MeV--23 GeV
%Experiments have been previously carried out for proton-activation of tellurium exists for proton
%energies 0.8--23 GeV
In these proton measurements, two long-lived radioisotopes were observed that have the 
potential to contribute background at the $0\nu\beta\beta$-decay peak:   
\textsuperscript{110m}Ag and \textsuperscript{60}Co.
%experiments have been performed using protons with energies 0.8--23 GeV
%point to 
Below 800 MeV, experimental data exists for activation of natural tellurium  
%no data exists for production of \textsuperscript{60}Co
%and \textsuperscript{110m}Ag by neutron or proton activation of tellurium. 
by $\sim$1--180 MeV neutrons \cite{Hansmann2010} and activation 
of individual tellurium isotopes by thermal to $\sim$15 MeV neutrons \cite{EXFOR}; 
however, only a few reactions were measured, and 
no cross sections were reported for the production of \textsuperscript{60}Co 
and \textsuperscript{110m}Ag.   
%this data includes cross sections for production 
%For the natural
%tellurium measurements, the neutrons range from $\sim$1--180 MeV \cite{HansmannThesis2010}.
%For the individual tellurium isotope measurements, the neutrons range
%from thermal to $\sim$15 MeV \cite{}.  However, none of these 
%reactions between neutrons and
%natural tellurium individual 
%stable tellurium isotopes and neutrons with energies up to approximately 18 MeV \cite{}.  
%This information can be used for neutron studies because at such high proton energies
%(i.e., much greater than the Coulomb barrier between a proton and a nucleus), 
%activation cross-sections for neutrons and protons are approximately equal.
%However, for energies between 20 MeV and 800 MeV, where much of the activation is
%expected to occur, no neutron-cross-section data exist.  
To deal with the lack of experimental data, the background from cosmogenic 
activation has been estimated in the past (as in Ref.~\cite{Lozza2014}) using a 
combination of the aforementioned neutron and proton measurements and codes 
that either implement the semi-empirical formulae by Silberberg and Tsao (S\&T)
\cite{Silberberg1973a,Silberberg1973b, Silberberg1998} 
(e.g., YIELDX \cite{Silberberg1973a,Silberberg1973b, Silberberg1998}, ACTIVIA \cite{Back2007}), or are based on 
Monte Carlo (MC) methods (e.g., CEM03 \cite{Mashnik2006}, HMS-ALICE \cite{Blann2006}, 
GEANT4 \cite{Agostinelli2003,Allison2006}).
%Most background analysis in this region relies on cross-sections that have been 
%calculated using semi-empirical models \cite{SilberbergTsao}.  
%For the available proton-activation data on tellurium, calculated and measured 
%cross-sections typically agree within about a factor of 2 
%\cite{Barghouty2013-ProtonsOnTe}\cite{}.  

These estimates can be greatly improved with additional neutron-activation 
cross-section measurements below 800 MeV, which can also be used to benchmark
the S\&T and MC codes. 
%to improve estimates of the cosmogenic activation background in 
%$0\nu\beta\beta$ experiments, as well as check the reliability of the  
%S\&T and MC codes.
%will therefore be needed to cover the intermediate 
%energy region, and the results can be used to further benchmark the calculations.
A sample of natural-TeO\textsubscript{2} powder 
was irradiated at the Los Alamos Neutron Science Center (LANSCE) with a
neutron beam containing neutrons with kinetic energies up to $\sim$800 MeV,
and having an energy distribution that resembles the cosmic-ray neutron flux at
sea-level.
%containing neutrons of up to $\sim$800 MeV kinetic energy 
%and having an energy distribution that resembles the cosmic-ray neutron flux at sea-level.
%
%that has an energy
%distribution that resembles the cosmic-ray neutron flux at sea-level. 
%
%containing neutron kinetic energies up to $\sim$800 MeV 
%and having an energy distribution that resembles the cosmic-ray neutron flux at sea-level.
%
%%in kinetic energy  beam containingthat resembles the cosmic-ray neutron flux at sea-level and contains 
%neutron energies up to 800 MeV. 
%neutron flux.  This beam contained neutrons with energies ranging from thermal up to 
%800 MeV. 
Following exposure, the $\gamma$ rays emitted from the sample were measured in a 
low-background environment with a high-purity-germanium (HPGe) detector to 
%identify
determine
the radioisotopes present.  Based on these results, flux-averaged cross sections 
%averaged over the 
%neutron flux 
were obtained for several dozen isotopes.  

The cross sections are used to investigate the impact cosmogenic activation 
will have on CUORE, a next-generation $0\nu\beta\beta$-decay experiment that 
will use an array of 988 high-resolution, low-background natural-TeO\textsubscript{2} 
bolometers to search for the $0\nu\beta\beta$ decay of \textsuperscript{130}Te.
In addition, the measured cross sections are compared with cross sections
calculated using the ACTIVIA code.
%available semi-empirical models.
Details of this measurement and subsequent
analysis are discussed below.

\section{Experimental method and data analysis}

\subsection{Target}
\label{sec:Target}
\begin{figure}[!tb]
   \includegraphics[width=0.5\textwidth]{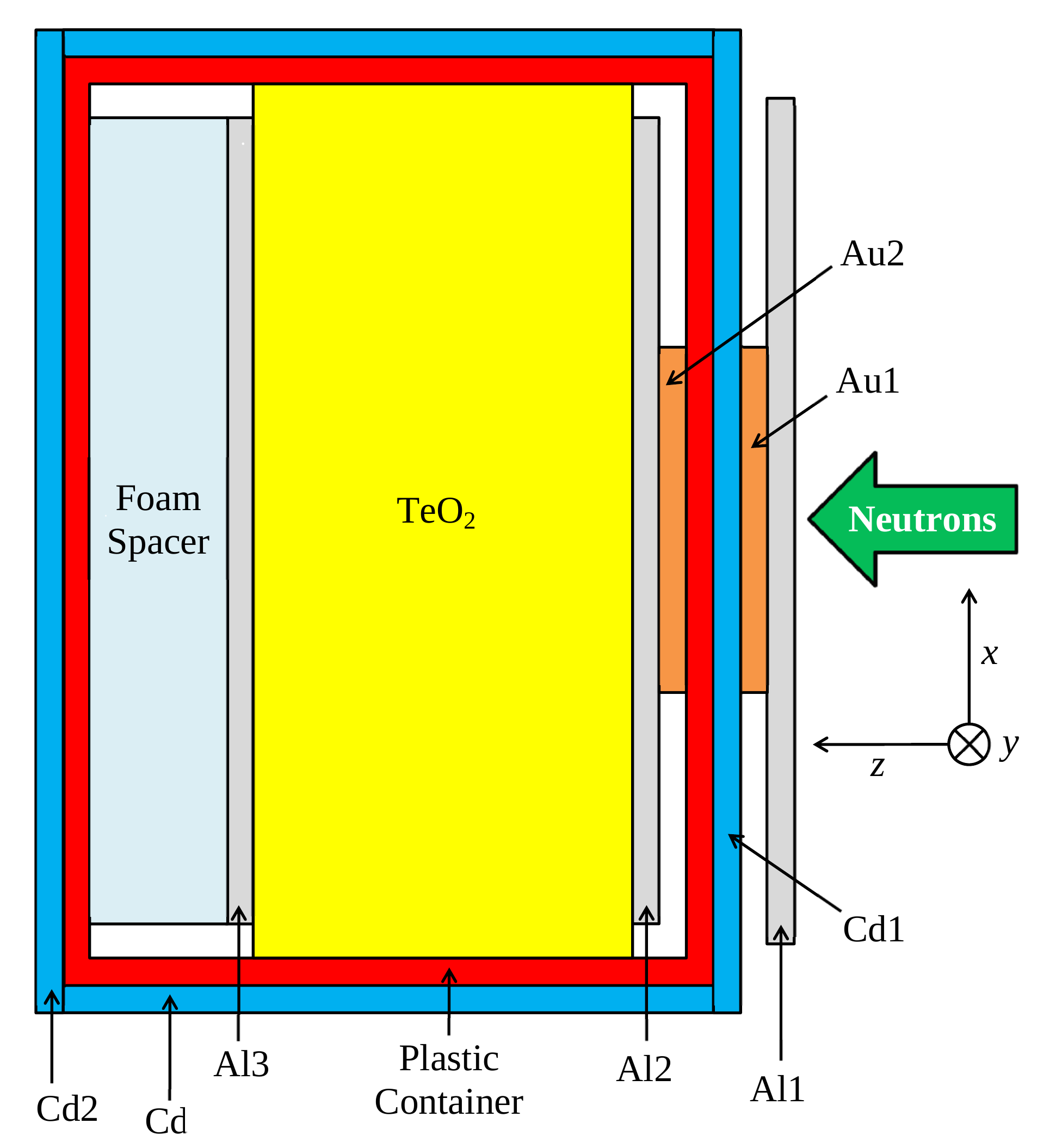}
   \caption
      {
      \label{fig:TargetSchematic}
      Schematic of the target irradiated at LANSCE.
      %; this drawing is not to scale.  
      The entire target is 6.2 cm long in the z direction.  
      Each target component has cylindrical symmetry about the z-axis.
      This drawing is not to scale.  
      %is a cylinder with its axis along the z-axis.
      Details on each component are given in Table~\ref{tab:TargetComponentTable}.
      }
\end{figure}

The target consisted of 272 g of natural-TeO\textsubscript{2} powder held within a 
cylindrical plastic container wrapped on all sides with 0.05 cm of 
cadmium to remove thermal neutrons.  The front and back cadmium-layers were 
also used to monitor the neutron flux on either side of the target.
Circular aluminum and gold foils were placed throughout the target 
%on either side of the target 
to monitor the neutron flux as well.  
%The plastic container was wrapped on all sides with a single layer of cadmium foil to remove 
%thermal neutrons and to monitor neutrons.  
The target geometry is illustrated in Figure~\ref{fig:TargetSchematic}, and the
details of each target component are listed in  
Table~\ref{tab:TargetComponentTable}. 
%contains details of each target component.

\begin{table*}
\caption{\label{tab:TargetComponentTable} Description of the target components illustrated 
in Figure~\ref{fig:TargetSchematic}.  The material, dimensions, mass, and purpose of each
component are given.  The parameters $\Delta z$ and $d$ are the thickness of 
the component along the z-axis and the diameter in the x-y plane, respectively.} 
\begin{ruledtabular}
\begin{tabular}{ l l c c c l }
  Component & Material & $\Delta z$ & $d$ & Mass & Purpose \\
   &  & (cm) & (cm) & (g) &  \\
  \hline\noalign{\smallskip}
  %\hline
  TeO\textsubscript{2} & TeO\textsubscript{2} powder & 2.79 & 6.43 & 271.56 & Target \\
  Al1 & Al & 0.0813 & 6.22 & 6.68 & Neutron-flux monitor \\
  Al2 & Al & 0.0813 & 5.93 & 6.06 & Neutron-flux monitor \\
  Al3 & Al & 0.0813 & 5.93 & 6.06 & Neutron-flux monitor \\
  Au1 & Au & 0.00515 & 2.54 & 0.504 & Neutron-flux monitor \\
  Au2 & Au & 0.00512 & 2.54 & 0.500 & Neutron-flux monitor \\
  Cd1 & Cd & 0.05 & 6.7 & 16.3 & Neutron-flux monitor \\
   & & & & & Thermal-neutron absorber \\
  Cd2 & Cd & 0.05 & 7.3 & 19.9 & Neutron-flux monitor  \\
   & & & & & Thermal-neutron absorber \\
  Cd & Cd & 0.05 & \textemdash & \textemdash & Thermal-neutron absorber \\
  Plastic Container & Polystyrene & 0.2 & \textemdash & \textemdash & Target holder \\
\end{tabular} 
\end{ruledtabular}
\end{table*}

\subsection{Neutron irradiation}
\label{sec:NeutronIrradiation}
\begin{figure}[!tb]
   \includegraphics[width=0.5\textwidth]{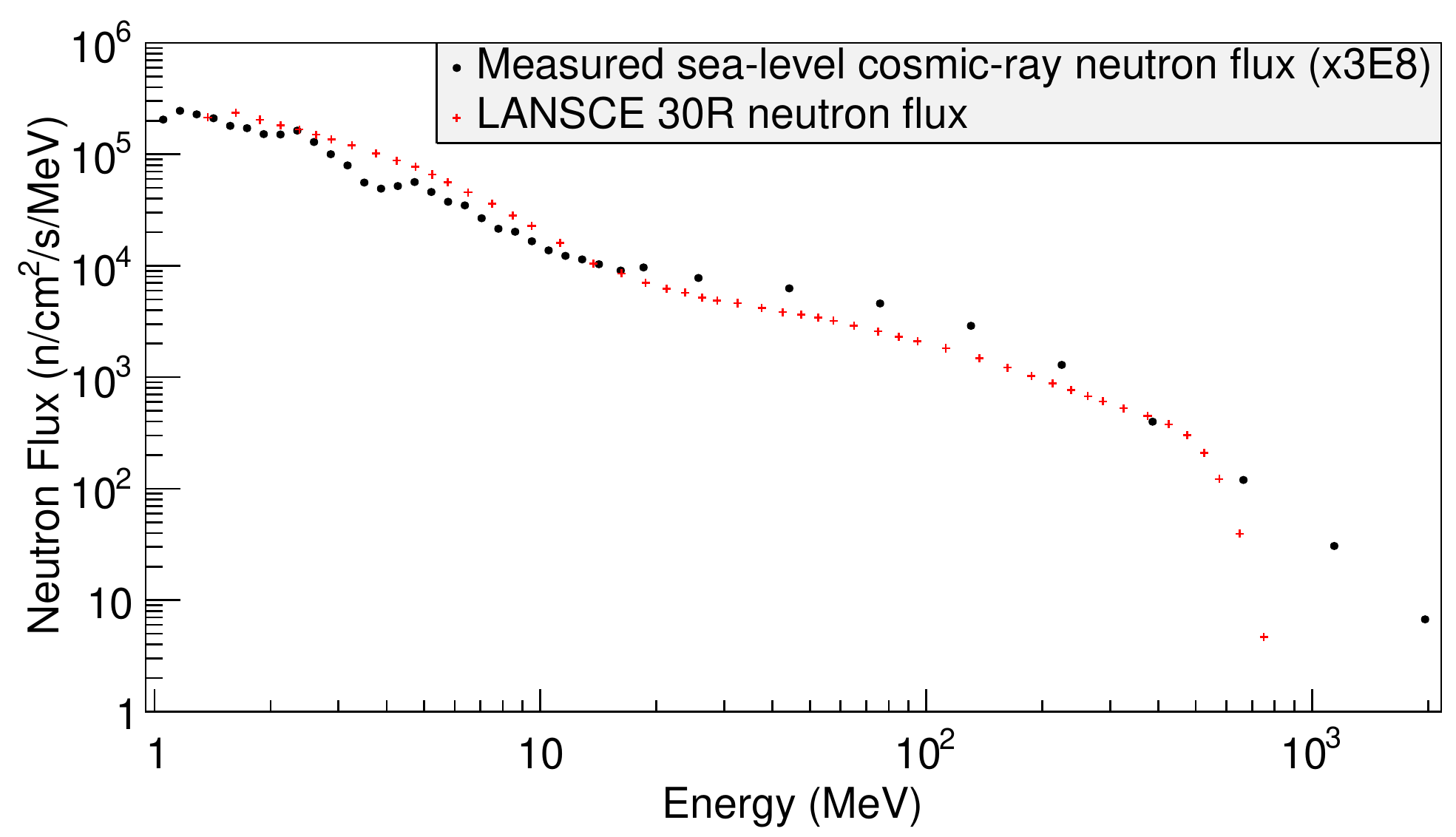}
   \caption
      {
      \label{fig:NeutronSpectrum}
      LANSCE 30R neutron flux (red) \cite{30RNeutronBeam} compared with the measured sea-level 
      cosmic-ray-neutron flux (black) \cite{Gordon2004}.  
      %(multiplied by a factor of $3\times10^{8}$).  
      }
\end{figure}

The target was irradiated with neutrons from the LANSCE 30R beam line for 43 hours 
during February 25--27, 2012.  
%The cross-section measurement was made by exposing a target sample to a  
%neutron beam at the LANSCE 
%Weapons Neutron Research (WNR) facility.  
At LANSCE, neutrons are generated from spallation 
reactions induced by an 800 MeV pulsed proton beam incident on a tungsten target.  
The 30R beam line, which is 30$\degree$ to the right of the proton beam, has
a neutron-energy spectrum that closely resembles the cosmic-ray neutron spectrum
at sea-level, but has an intensity $3\times 10^8$ times larger, 
%The neutron beam line 30$\degree$ to the right (30R) of the proton beam was chosen for this 
%experiment because it closely resembles the cosmic-ray-neutron spectrum at sea-level,
%but has an intensity $3\times 10^8$ times larger. 
%The 30R neutron spectrum is 
as shown in Figure~\ref{fig:NeutronSpectrum}.  
%and compared with the measured cosmic-ray-neutron spectrum at sea-level \cite{Gordon2004-CRNeutronSpectrum}. 
%(multiplied by $3\times10^{8}$).
A beam collimation width of 8.26 cm was used, which resulted in a beam-spot diameter
of 8.41 cm at the target. 

The proton beam used to generate the neutrons consisted of 625-$\mu$s-long 
macropulses occuring at a rate of 40 Hz.  Each macropulse contained micropulses 
spaced 1.8 $\mu$s apart. 
The neutron time-of-flight was obtained by measuring the time between 
the arrival of the proton macropulse at the tungsten target and the 
generation of a fission signal in a \textsuperscript{238}U-fission ionization 
chamber \cite{Wender1993} located 25.4 cm upstream of the TeO\textsubscript{2} 
target.  
%The flux of neutrons impinging on the TeO\textsubscript{2} target 
%was measured using time-of-flight and a \textsuperscript{238}U-fission 
%ionization chamber \cite{Wender1993}
%located 25.4 cm upstream of the TeO\textsubscript{2} target.  
%The proton beam used to generate the neutrons 
%consisted of 625-$\mu$s-long macropulses occuring at a rate of 40 Hz.  Each
%macropulse contained micropulses spaced 1.8 $\mu$s apart. 
%The velocity, and 
%therefore energy, of a neutron is obtained by measuring the time between 
%the arrival of the proton macropulse at the tungsten target and the 
%generation of a fission signal in the ionization chamber.  
The ionization chamber was only able to detect neutrons with energies above
the \textsuperscript{238}U-fission threshold, which is approximately 1.25 MeV. 
%Since fission of \textsuperscript{238}U is negligible at neutron energies 
%below approximately 1.25 MeV, the ionization chamber is only able to record 
%neutrons with energies above 1.25 MeV.  
The average neutron flux above 1.25 MeV at the TeO\textsubscript{2} target was 
determined to be $1.41 \times 10^6$ neutrons/(cm\textsuperscript{2}$\cdot$s), 
with an estimated uncertainty of 10\% \cite{Wang2014} 
based on uncertainties in the geometry and efficiency of the ionization chamber.
%; this takes into account uncertainties in 
%the beam-spot size, the flight-path distance, the isotopic composition of the fission 
%chamber foils, the \textsuperscript{238}U fission cross-sections, etc.  

\subsection{Gamma-ray analysis of the irradiated target}
\label{sec:GammaCountingTheIrradiatedTarget}
\begin{figure}[!htbp]
    \centering
%    \begin{subfigure}[t]{0.5\textwidth}
%       \centering
%       \includegraphics[width=\textwidth]{GoldFoilCountingSetup.pdf}
%       \caption{Setup used to count each gold foil.}
%       \label{fig:GoldFoilCountingSetup}
%    \end{subfigure}
%    \quad
%    \begin{subfigure}[t]{0.5\textwidth}
%       \centering
%       \includegraphics[width=\textwidth]{MonitorFoilCountingSetup.pdf}
%       \caption{Setup used to count each cadmium and aluminum foil.}
%       \label{fig:MonitorFoilCountingSetup}
%    \end{subfigure}
%    \quad
%    \begin{subfigure}[t]{0.5\textwidth}
%       \centering
       \includegraphics[width=0.5\textwidth]{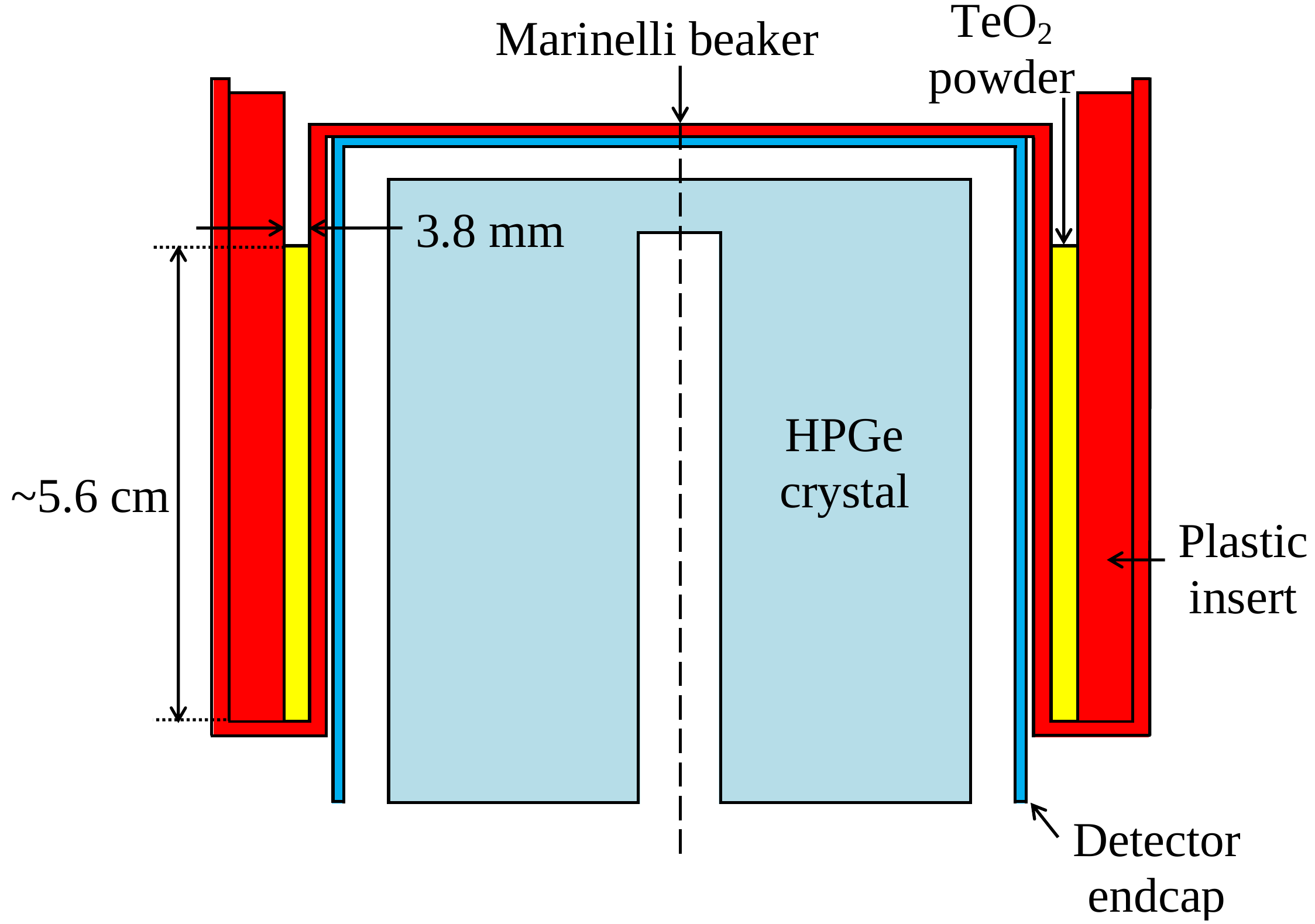}
%       \caption{Setup used to count TeO\textsubscript{2} powder.}
%       \label{fig:TeO2PowderCountingSetup}
%    \end{subfigure}
%    \quad
    \caption
       {
       \label{fig:TeO2PowderCountingSetup}
%       \label{fig:CountingSetups}
       Setup used during the $\gamma$-ray measurement of the      
       %to gamma-count the 
       TeO\textsubscript{2} powder.  Each component has cylindrical symmetry 
       about the dashed line.  This drawing is not to scale.
       }
\end{figure}

\begin{figure}[!htbp]
   \centering
   %\begin{subfigure}[t]{0.5\textwidth}
   %   \centering
   %   \includegraphics[width=\textwidth]{EnergySpectrum_28229S_NewEnergyCalibration.pdf}
      %\caption{Complete spectrum.  }
   %   \label{fig:TeO2GammaSpectra_a}
   %\end{subfigure}
   %\qquad
   %\begin{subfigure}[t]{0.5\textwidth}
   %   \centering
   %   \includegraphics[width=\textwidth]{EnergySpectrum_28229S_NewEnergyCalibration_650-900keV.pdf}
      %\caption{Zoomed-in view of spectrum with peaks of interest labeled.}
   %   \label{fig:TeO2GammaSpectra_b}
   %\end{subfigure}
   \includegraphics[width=0.5\textwidth]{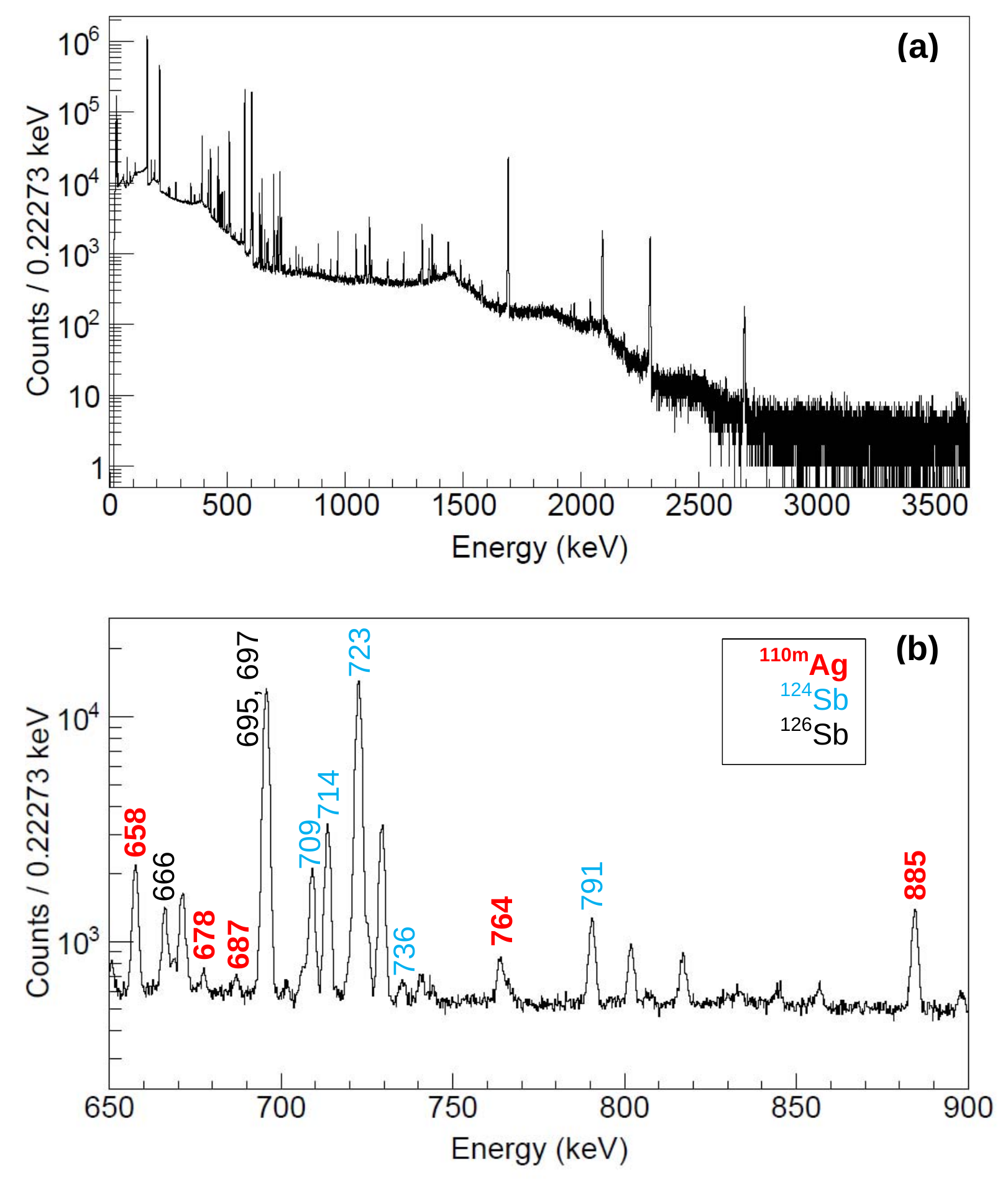}
   \caption
      {
      \label{fig:TeO2GammaSpectra}
      3-day-long $\gamma$-ray spectrum collected for the  
      TeO\textsubscript{2} powder four months after the neutron irradiation.  
      (a) Full spectrum.
      (b) A region of the spectrum where \textsuperscript{110m}Ag 
      peaks were observed.  Labeled peaks are associated with the decay of 
      isotopes with Q values greater than the Q\textsubscript{$\beta\beta$}
      of \textsuperscript{130}Te, i.e., \textsuperscript{110m}Ag (red and bolded),
      \textsuperscript{124}Sb (blue), and \textsuperscript{126}Sb (black). 
      Other peaks in the region are from the decays of 
      \textsuperscript{125}Sb, \textsuperscript{129m}Te, 
      \textsuperscript{105}Ag, and \textsuperscript{114m}In.
      }
\end{figure}

Approximately one week after the neutron irradiation, the TeO\textsubscript{2} 
target was dismantled, and each component was analyzed using $\gamma$-ray 
spectroscopy at the 
%gamma-counted with an upright, 115\%, n-type HPGe detector at the 
Lawrence Berkeley National Laboratory %(LBNL) 
Low Background Facility \cite{Thomas2013_2,BLBF}. 
%(LBF).  
%To identify the radioisotopes produced by neutron activation, 
%each TeO\textsubscript{2} target component was gamma counted at the LBF.  
The TeO\textsubscript{2} powder, cadmium foils, and aluminum foils were 
measured using an upright, $115 \%$-relative-efficiency, n-type HPGe detector, and 
the gold foils were measured with a horizontal, $80 \%$-relative-efficiency, 
p-type HPGe detector.
Each detector was surrounded by a copper inner shield encased in a lead
outer shield.
%Both detectors were encased in lead housings.
The gold foils were highly activated and could be counted
%far away from the detector.  
at a distance of 
12 cm from %the center of 
the detector. %face was chosen  
%to simplify the detection-efficiency calibration. 
%The gold foils were highly activated and were each counted 12 cm away from the 
%front of the detector to simplify the detection-efficiency calibration. 
%minimize the effects of true-coincidence summing and
%Their configuration is shown in Figure~\ref{fig:GoldFoilCountingSetup}.     
%minimize counting time.
%that required each to be measured close to the detectorthem to be measured close to the detectormeasuring them close to the detector to 
%maximize counting statistics and minimize counting time
%and were each counted 
%The cadmium and aluminum monitor foils had low levels of activity and 
The cadmium and aluminum foils had low 
levels of activity and were therefore measured directly on top of the detector to 
%close to the detector to
maximize the detection efficiency. 
%For the cadmium and aluminum measurements, 
%each foil was laid directly on top of the detector.
%in the configuration illustrated in Figure~\ref{fig:MonitorFoilCountingSetup}, 
%with the foil laid flat and centered on top of the detector to maximize
%counting statistics and minimize the counting time.  
%Prior to counting, 
For the TeO\textsubscript{2} powder, the $\gamma$-ray measurements needed to be
highly sensitive to long-lived radioisotopes, which had low levels of activity
inside the powder.  To maximize the detection efficiency,  
%For the TeO\textsubscript{2} measurements, the irradiated 
the TeO\textsubscript{2} powder was mixed thoroughly and counted in a Marinelli 
beaker positioned over the top of the detector 
(Figure~\ref{fig:TeO2PowderCountingSetup}).  
A plastic insert was placed
%glued to the outer part of the 
inside the beaker to decrease the thickness 
and increase the height of the powder, which in turn 
%This geometry, illustrated in Figure~\ref{fig:TeO2PowderCountingSetup},
increased the solid angle of the detector seen by the powder and 
decreased the self-attenuation of $\gamma$ rays from decays within the powder. 
The thickness and average height of the TeO\textsubscript{2} powder were 3.8 mm and 
$\sim$5.6 cm, respectively.
%because it increases the probability, or efficiency, 
%for measuring gamma rays by minimizing the self-attenuation of gamma rays
%through the powder and by maximizing the solid angle of the detector seen
%by the powder. 
The TeO\textsubscript{2} was counted in this configuration periodically for
six months to enable the observation of long-lived activation products 
after the short-lived ones decayed away.  
Figure~\ref{fig:TeO2GammaSpectra} 
shows a $\gamma$-ray spectrum for the TeO\textsubscript{2} 
powder collected four months after the irradiation. 
%The numerous peaks in the top spectrum come from the decays of 
%isotopes with half-lives as short as 1.39 d 
%(\textsuperscript{131m}Te).  In the bottom spectrum, fewer lines are 
%present because the short-lived activation products have decayed away,
%leaving only products with half-lives greater than $\sim$12 d.  

Each peak in the $\gamma$-ray spectra was fit with 
a Gaussian summed with a quadratic background function
to determine the energy and net counts.
For peaks with higher intensity,  
%a skewed Gaussian and 
a smoothed step function was also added to 
the fitting function. 
%to describe incomplete charge collection and escape of photoelectrons from the detector. 
%the sum of a Gaussian, skewed Gaussian, smoothed step function, and
%quadratic function 
%either a Gaussian or a Gaussian-like function summed with a quadratic 
%function representing the Compton continuum. 
%The peak-fitting was performed with the software RadWare \cite{RadWare}.
The $\gamma$-ray energies were used to identify
the radioisotopes produced in the TeO\textsubscript{2} powder.  
For $\gamma$-ray lines that could come from the decay of more than one isotope,
the contributors were identified from the decay half-life of the line. 

A list of the radioisotopes observed in the TeO\textsubscript{2} powder is
provided in Table~\ref{tab:RadioisotopesObservedInTeO2}.  
Since $\gamma$-ray measurements started one week after the neutron irradiation ended,
only activated isotopes with half-lives greater than 
$\sim$1 day remained.
Therefore, any observed isotope 
%present in the TeO\textsubscript{2} powder that has 
%a half-life less than $\sim$1 d 
with a shorter half-life was a decay daughter of a longer-lived isotope.  
For example, the presence of \textsuperscript{127}Te (9.35-hour half-life) and 
\textsuperscript{129}Te (69.6-minute half-life) 
%with half-lives of ... and ..., respectively, 
was due to the decays of the longer-lived metastable states 
\textsuperscript{127m}Te and \textsuperscript{129m}Te, respectively.

%For certain isotopes, multiple production paths were possible. 
%\textsuperscript{127m}Te, for example, could be directly produced from the reactions 
%\textsuperscript{126}Te(n,$\gamma$), \textsuperscript{128}Te(n,2n), and 
%\textsuperscript{130}Te(n,4n).  It could also be indirectly produced from the decay
%of \textsuperscript{127}Sb, another isotope activated in the TeO\textsubscript{2} powder. 
%In column two of Table~\ref{tab:RadioisotopesObservedInTeO2}, Te(n,X) or O(n,X) are used 
%to indicate that the isotope's presence in the TeO\textsubscript{2} powder 
%is due to more than one type of reaction with tellurium or oxygen 
%during the irradiation.  
%The isotope \textsuperscript{131m}Te is an exception in that it could only 
%be produced via \textsuperscript{130}Te(n,$\gamma$); this specific reaction %is therefore reported in the table. 

%\begin{longtable}[c]{+l^c^c^r^c^c}
%\begin{table*}[c]{+l^c^c^r^c^c}
\begin{table*}
\caption{\label{tab:RadioisotopesObservedInTeO2} Radioisotopes observed in
the irradiated TeO\textsubscript{2} powder.
Unless otherwise indicated, all isotopes were produced by neutron
interactions with tellurium.
%The minimum threshold energies for these interactions are given in column 
%three for isotopes with half-lives greater than 1 d.  
%Te(n,X) reactions 
%producing the isotope are given in column three for isotopes with
%half-lives greater than 1 d.  
%No threshold energy is provided for isotopes
%that could be produced by Te(p,X) reactions. 
The measured and calculated flux-averaged cross sections  
($\bar{\sigma}_{\text{30R}}$ and $\bar{\sigma}$\textsubscript{S\&T}, respectively)
for neutron activation of tellurium are provided %in columns 
%four and five 
for isotopes with half-lives greater than 1 day.  All $\bar{\sigma}_{\text{30R}}$
%measured cross sections 
were measured at the $68 \%$ C.L. and
are independent cross sections, except for those followed by ``(cu)," which are cumulative.  
All rows corresponding to isotopes that can contribute background at the
\textsuperscript{130}Te $0\nu\beta\beta$-decay peak are bolded, and for these 
isotopes, the decay modes ($\epsilon$ and $\beta^-$ for electron capture and 
beta-minus decay, respectively) and Q values are given.
%No cross sections were obtained for isotopes that were produced solely by indirect methods.
}
\begin{ruledtabular}
\begin{tabular}{+l^r^c^c^c}
   \multicolumn{1}{c}{Isotope} & \multicolumn{1}{r}{Half-life} \comment{& \multicolumn{1}{c}{Minimum Threshold Energy}} & \multicolumn{1}{c}{$\bar{\sigma}_{\text{30R}}$ ($68 \%$ C.L.)} & \multicolumn{1}{c}{$\bar{\sigma}$\textsubscript{S\&T}} & \multicolumn{1}{c}{Decay Q value} \\
   \multicolumn{1}{c}{} & \multicolumn{1}{r}{} \comment{& \multicolumn{1}{c}{(MeV)}} & \multicolumn{1}{c}{(mb)} & \multicolumn{1}{c}{(mb)} & \multicolumn{1}{c}{(MeV)}\\
   \hline\noalign{\smallskip}
   \textsuperscript{126}I\footnote{\label{ProtonInteractionNote}This 
isotope was produced by interactions with 
spallation protons created in the target during the neutron irradiation.
%proton interactions with TeO\textsubscript{2} 
%during the neutron irradiation.  These protons were generated from neutron
%spallation reactions with materials in the target.  
Therefore, no \comment{flux-averaged} cross sections are provided.}  & 12.93 d \comment{&}  &  & &\\
   \textsuperscript{131}I\footref{ProtonInteractionNote} & 8.025 d \comment{&}  &  & &\\
   \textsuperscript{118}Te & 6.00 d \comment{& 18.0} & $5.7 \pm 1.2$ & 9.80 &\\
   \rowstyle{\bfseries}
   \textsuperscript{119m}Te & 4.7 d \comment{& 10.6} & $6.3 \pm 0.8$ &13.0 & 2.554 ($\epsilon$) \\
   \textsuperscript{121}Te\footnote{\label{LowEnergyNeutronNote}This isotope
had a high probability of being produced by interactions with $< 1.25$-MeV neutrons.
Therefore, no \comment{flux-averaged} cross sections are given.} & 19.17 d \comment{& 0.0} &  & &\\
   \textsuperscript{121m}Te & 164.2 d \comment{& 0.0} & $16 \pm 2$ & 25.2 &\\
   \textsuperscript{123m}Te & 119.2 d \comment{& 0.0} & $36 \pm 4$ & 8.6&\\
   \textsuperscript{125m}Te & 57.4 d \comment{& 0.0} & $83 \pm 10$ &17.0 &\\
   \textsuperscript{127}Te & 9.35 h \comment{&}  & & &\\
   \textsuperscript{127m}Te & 106.1 d \comment{& 0.0} & $46 \pm 9$ &25.3 &\\
   \textsuperscript{129}Te & 69.6 m \comment{&}  & & &\\
   \textsuperscript{129m}Te & 33.6 d \comment{& 0.0} & $53 \pm 17$ (cu)&22.4 (cu)&\\
   \textsuperscript{131}Te & 25 m \comment{&}  & & &\\
   \textsuperscript{131m}Te\footref{LowEnergyNeutronNote} & 33.25 h \comment{& 0.0} & & &\\
   \textsuperscript{131m}Xe\footref{ProtonInteractionNote} & 11.84 d \comment{&}  & & &\\
   \bfseries{\textsuperscript{118}Sb} & \bfseries{3.6 m} \comment{&}  & & & \bfseries{3.657 ($\epsilon$)} \\
   \textsuperscript{119}Sb\footnote{A flux-averaged cross section could 
not be obtained for \textsuperscript{119}Sb because the strongest 
$\gamma$-ray line at 24 keV overlapped with x-rays emitted 
by other activated isotopes.} & 38.19 h \comment{& 5.0} &  & &\\
   \rowstyle{\bfseries}
   \textsuperscript{120m}Sb & 5.76 d \comment{& $> 0.2$} & $6.3 \pm 0.8$ &10.2 & $2.681+E_{ex}$ ($\epsilon$) \\
   \textsuperscript{122}Sb & 2.7238 d \comment{& 1.2} & $14 \pm 2$ (cu) &15.4 (cu)&\\
   \rowstyle{\bfseries}
   \textsuperscript{124}Sb & 60.2 d \comment{& 2.1} & $16 \pm 2$ (cu)&19.1 (cu)& 2.904 ($\beta^{-}$) \\
   \textsuperscript{125}Sb & 2.759 y \comment{& 0.0} & $18 \pm 2$ (cu) &18.8 (cu)&\\
   \rowstyle{\bfseries}
   \textsuperscript{126}Sb & 12.35 d \comment{& 2.9} & $6.7 \pm 0.9$ (cu) &26.4 (cu)& 3.673 ($\beta^{-}$) \\
   \textsuperscript{127}Sb & 3.85 d \comment{& 7.4} & $13 \pm 2$ (cu) &9.8 (cu)&\\
   \textsuperscript{113}Sn & 115.1 d \comment{& 27.9} & $2.6 \pm 0.3$ (cu) &3.0 (cu)&\\
   \textsuperscript{117m}Sn & 14 d \comment{& 0.0} & $4.3 \pm 0.6$ &0.63 &\\
   \textsuperscript{111}In & 2.805 d \comment{& 34.8} & $2.3 \pm 0.3$ (cu) &2.1 (cu)&\\
   \textsuperscript{114m}In & 49.51 d \comment{& 10.4} & $1.9 \pm 0.2$ &0.31 &\\
   \textsuperscript{105}Ag & 41.29 d \comment{& 77.7} & $0.56 \pm 0.07$ (cu) &0.45 (cu)&\\
   \rowstyle{\bfseries}
   \textsuperscript{106m}Ag & 8.28 d \comment{& 58.4} & $0.44 \pm 0.09$ &0.39 & 3.055 ($\epsilon$) \\
   \bfseries{\textsuperscript{110}Ag} & \bfseries{24.56 s} \comment{&}  &  & & \bfseries{2.893 ($\beta^{-}$)} \\
   \rowstyle{\bfseries}
   \textsuperscript{110m}Ag & 249.83 d \comment{& 13.9} & $0.28 \pm 0.04$ & 0.054 & 3.010 ($\beta^{-}$) \\
   \textsuperscript{111}Ag & 7.45 d \comment{& 11.2} & $0.42 \pm 0.09$ (cu) &0.030 (cu)&\\
   \textsuperscript{101}Rh & 3.3 y \comment{& 80.0} & $0.06 \pm 0.01$ (cu) &0.24 (cu)&\\
   \textsuperscript{101m}Rh & 4.34 d \comment{& 80.0} & $0.30 \pm 0.05$ (cu) &0.24 (cu)&\\
   \textsuperscript{102m}Rh & 3.742 y \comment{& 80.0} & $0.15 \pm 0.02$ &0.12 &\\
   \rowstyle{\bfseries}
   \textsuperscript{60}Co\footnote{\textsuperscript{60}Co was not conclusively
observed in the $\gamma$-ray spectra due to \textsuperscript{102m}Rh and
\textsuperscript{110m}Ag peaks being present where the \textsuperscript{60}Co 
peaks were expected.  Therefore the cross section quoted for 
\textsuperscript{60}Co is an upper limit.} & 5.27 y \comment{& 80.0} & $< 0.0016$ (cu) &0.0013 (cu) & 2.823 ($\beta^{-}$) \\
   \textsuperscript{7}Be\footnote{\textsuperscript{7}Be was produced 
almost exclusively by neutron interactions with oxygen.  The 
cross sections given correspond to these interactions.} & 53.24 d \comment{& 35.5} & $1.4 \pm 0.2$ & 2.5&\\
\end{tabular}
\end{ruledtabular}
\end{table*}

\subsection{Photopeak efficiencies}
\label{sec:PhotopeakEfficiencies}
The $\gamma$-ray measurements of the TeO\textsubscript{2} powder needed to be
highly sensitive to long-lived radioisotopes, which had low levels of activity
inside the powder.  To maximize the detection efficiency,  
%The TeO\textsubscript{2} powder contained low levels of activity, making it
%necessary to count it close to the detector during the $\gamma$-ray 
%measurements in order to maximize the detection efficiency 
the powder was counted immediately next to the detector (Figure~\ref{fig:TeO2PowderCountingSetup}). 
Determination of the photopeak efficiencies for the 
TeO\textsubscript{2} powder from calibration measurements alone was impractical 
due to the complexity of the counting geometry and the effects of 
true-coincidence summing, which can be significant at such close range. 
%depending on
%the decay scheme of the isotope and the total efficiencies of the $\gamma$ 
%rays in coincidence.
Therefore, the efficiencies were obtained by running 
simulations with the Geometry and Tracking 4 (GEANT4) code, version 4.9.4.p02, which were 
%The simulations 
benchmarked against experimental measurements of various point and extended 
$\gamma$-ray sources (Table~\ref{tab:Sources}) that 
covered a wide range of $\gamma$-ray energies.
%that had varying complexities in geometry and 
%in a variety of different locations and extended geometries that
%covered a wide range of $\gamma$-ray energies.

%, experimental determination of 
%photopeak efficiencies that take into account summing is impractical.  

%At such close range, true-coincidence summing can have a significant impact on 
%the photopeak efficiencies, depending on the decay scheme of the isotope     
%and the total efficiencies of the $\gamma$ rays in coincidence.
 
%Prior to determining the photopeak efficiencies, 
%The photopeak efficiencies determined using GEANT4 were benchmarked 
%against experimental measurements of $\gamma$-ray sources 
%(Table~\ref{tab:Sources}) that had varying complexities in geometry and 
%covered a wide range of $\gamma$-ray energies.
%Two point sources, \textsuperscript{57}Co and \textsuperscript{54}Mn;
%a uranium source consisting of a uranium-ore and epoxy mixture spread
%inside a cylindrical plastic container; and two extended sources, each of which
%consisted of a powder mixture inside a Marinelli beaker, were used. 
%Note that all Marinelli beakers used in the benchmarking measurements
%and measurements of the TeO\textsubscript{2} powder are the same, with
%identical plastic inserts glued to the inside. 
For the benchmarking measurements, the \textsuperscript{57}Co and 
\textsuperscript{54}Mn point sources were each 
counted at the center of the detector face and at four positions along the side of 
the detector that were spaced 2 cm apart and spanned the length of the 
HPGe crystal.  The uranium source was counted %while taped to the outside of a Marinelli beaker.  
on the side of the detector as well.  Following the natural-source 
method \cite{PerilloIsaac1997}, the two extended sources, ES1 and ES2, 
were constructed from powders that contained elements with naturally-occurring 
long-lived radioisotopes. 
ES1 was designed to mimic the geometry of the irradiated TeO\textsubscript{2}
powder during the $\gamma$-ray measurements, and ES2 was designed to mimic
both the geometry and density of the powder.
%designed to mimic the geometry and/or density of the irradiated TeO\textsubscript{2} 
%powder during the $\gamma$-ray measurements.  
Photopeak efficiencies were obtained for all the $\gamma$ rays listed in
Table~\ref{tab:Sources}.  In addition, the total efficiency, which is needed to
determine summing corrections, was obtained for the two \textsuperscript{57}Co
$\gamma$ rays (122.06 keV and 136.47 keV) and the \textsuperscript{54}Mn 
$\gamma$ ray (834.85-keV). 

\begin{table*}
\centering
\caption{\label{tab:Sources} Description of $\gamma$-ray sources used to 
benchmark GEANT4.
}
\begin{ruledtabular}
\begin{tabular}{l l l c c}
%   \hline\noalign{\smallskip}
   Source & Composition & Dimensions & $\gamma$-ray & Branching Ratio \\
    &  &  & (keV) & (\%) \\
   \hline\noalign{\smallskip}
   Co-57 & Co-57 & Point source & 122.06 & $85.60 \pm 0.17$ \\
    & & & 136.47 & $10.68 \pm 0.08$  \\
   Mn-54 & Mn-54 & Point source & 834.85 & $99.9760 \pm 0.0010$ \\
   Uranium\footnote{All isotopes in the source were assumed to be in secular equilibrium.} & Natural uranium ore (0.1176 g) & Diameter = 4.76 cm & 185.72 (\textsuperscript{235}U) & $57.2 \pm 0.8$ \\
    & mixed with epoxy & Thickness = 3.175 mm & 46.54 (\textsuperscript{210}Pb) & $4.25 \pm 0.04$ \\
    & & & 186.21 (\textsuperscript{226}Ra) & $3.64 \pm 0.04$ \\
    & & & 242.00 (\textsuperscript{214}Pb) & $7.251 \pm 0.016$ \\
    & & & 295.22 (\textsuperscript{214}Pb) & $18.42 \pm 0.04$ \\
    & & & 1764.49 (\textsuperscript{214}Bi) & $15.30 \pm 0.03$ \\
    & & & 2204.06 (\textsuperscript{214}Bi) & $4.924 \pm 0.018$ \\
   ES1\footnote{\label{Ac227Note}Due to a small \textsuperscript{227}Ac contamination in the 
La\textsubscript{2}O\textsubscript{3}, ES1 and ES2 also contained 
\textsuperscript{227}Ac and its daughter isotopes, which were assumed to
be in secular equilibrium with each other.  $\gamma$-rays from the
\textsuperscript{227}Ac chain are also listed in the table.}  
      & La\textsubscript{2}O\textsubscript{3} powder (89 g), & Inner radius = 5.06 cm & 201.83 (\textsuperscript{176}Lu) & $78.0 \pm 2.5$  \\
    & Lu\textsubscript{2}O\textsubscript{3} powder (2 g), & Outer radius = 5.443 cm & 306.78 (\textsuperscript{176}Lu) & $93.6 \pm 1.7$ \\
    & KCl powder (4 g) & Average height = 5.75 cm & 788.74 (\textsuperscript{138}La) & $34.4 \pm 0.5$ \\
    & & & 1435.80 (\textsuperscript{138}La) & $65.6 \pm 0.5$ \\
    & & & 1460.82 (\textsuperscript{40}K)  & $10.66 \pm 0.18$ \\
    & & & 269.46 (\textsuperscript{223}Ra) & $13.9 \pm 0.3$ \\
    & & & 271.23 (\textsuperscript{219}Rn) & $10.8 \pm 0.6$ \\
    & & & 832.01 (\textsuperscript{211}Pb) & $3.52 \pm 0.06$ \\
    & & & 351.07 (\textsuperscript{211}Bi) & $13.02 \pm 0.12$ \\
   ES2\footref{Ac227Note} 
      & (Unirradiated) TeO\textsubscript{2} powder (228 g), & Inner radius = 5.06 cm & Note: All $\gamma$ rays used &\\
    & La\textsubscript{2}O\textsubscript{3} powder (23 g), & Outer radius = 5.443 cm & to analyze ES1 were also &\\
    & Lu\textsubscript{2}O\textsubscript{3} powder (6 g), & Average height = 6.5 cm & used to analyze ES2. &\\
%      & (Unirradiated) TeO\textsubscript{2} powder (228 g), & Inner radius = 5.06 cm & \multicolumn{1}{l}{Note: All $\gamma$ rays used} &\\
%    & La\textsubscript{2}O\textsubscript{3} powder (23 g), & Outer radius = 5.443 cm & \multicolumn{1}{l}{to analyze ES1 were also} &\\
%    & Lu\textsubscript{2}O\textsubscript{3} powder (6 g), & Average height = 6.5 cm & \multicolumn{1}{l}{used to analyze ES2.} &\\
    & K\textsubscript{2}SO\textsubscript{4} powder (14 g) &  & \\
%    \hline
\end{tabular}
\end{ruledtabular}
\end{table*}

%For both ES1 and ES2, the 202-keV, 307-keV, and 1436-keV gamma rays all 
%suffered from the effects of summing out.  
%GEANT4 is a C++-based software package comprised of tools that can be used
%to accurately simulate the passage of particles through matter.
%Before running the code, the user must: 
%\begin{itemize}
%  \item construct the simulated system by defining the geometry and 
%  material of each object in the system (e.g., the detector, items 
%  surrounding the detector); 
%  \item indicate what physics processes (e.g., Compton scattering, 
%  photoelectric effect, ionization, bremsstrahlung) 
%  will be included in the simulation to describe how particles interact 
%  with materials;
%  \item construct the radiation source by defining its geometry, its 
%  location, the particles it emits, etc.    
%\end{itemize}
%Once the simulation is started, GEANT4 utilizes Monte Carlo methods 
%\cite{Robert2004} to track the source particles as they travel through the 
%system.  
%The user can ask GEANT4 to output information 
%such as the energy deposition in specified parts of the system, 
%track lengths of source particles, physics processes the source
%particles participate in, the secondary particles produced in each 
%interaction, etc.

The benchmarking measurements were simulated using
GEANT4.  Each simulation included the HPGe detector, the $\gamma$-ray
source, and the lead and copper shielding. %, all of which were constructed
%in GEANT4 using the materials and geometries specified by their
%manufacturers or given in Table~\ref{tab:Sources}.  
%For the extended sources, the height of the powder in the simulation 
%was taken to be the average height of the powder during the measurement. 
%Photopeak efficiencies were determined by first constructing the
%gamma-counting setup for the TeO\textsubscript{2} powder 
%(Figure~\ref{fig:TeO2PowderCountingSetup}) in GEANT4.  The setup included
%All components in the simulation were constructed using the materials and
%geometries specified by their manufacturers.  
%A separate simulation was performed 
For each $\gamma$ ray of interest,  
%during which 
the entire decay scheme of the parent nucleus was simulated. %taken into account.
Angular correlations between coincident $\gamma$ rays  
were not taken into account; however at close distances
to the detector, the effects on the photopeak efficiencies are largely 
averaged out and are thus small.  
%For $\gamma$ rays with negligible or no summing, monoenergetic, isotropic
%$\gamma$-ray sources were constructed in GEANT4.  
%For $\gamma$ rays with noticeable summing (201.83 keV, 306.78 keV, 1435.80 
%keV), the parent isotope was simulated.
%In the uranium-ore and extended source simulations, the $\gamma$ rays or
%isotopes were generated uniformly throughout the source volume. 
%Enough events were run so that the statistical uncertainty in the counts
%for each peak was less than 1\%.

Each simulated photopeak or total efficiency ($\epsilon_{s\gamma}$) was 
compared with the measured value ($\epsilon_{m\gamma}$), and the percent difference 
was determined:

\begin{equation}
   \Delta\epsilon_{\gamma} = \frac{\epsilon_{m\gamma} - \epsilon_{s\gamma}}{\epsilon_{s\gamma}}\times100\%.
\end{equation}
%Here, $\epsilon_{m\gamma}$ and $\epsilon_{s\gamma}$ are the measured and
%simulated photopeak (or total) efficiencies, respectively.
Using the manufacturer's detector specifications in the  
simulations resulted in $\Delta\epsilon_{\gamma}$ values 
%for the photopeak efficiencies 
that ranged from approximately -10\% to -35\%,  
with the agreement between simulation and measurement 
worsening at lower $\gamma$-ray energies.
This kind of disagreement, especially overestimation by the simulation, has 
been seen in other studies that 
%also implemented Monte Carlo methods to 
model the $\gamma$-ray efficiencies of HPGe detectors using the geometry
provided by the manufacturer (e.g., Refs.~\cite{Boson2008, Huy2007, Helmer2003, Wang2002}).
Typically, the discrepancies have been attributed to physical characteristics of
the detector (crystal location, Li-diffused-contact thickness, etc.) that
are difficult for the manufacturer to precisely specify.  
When the source is counted close to the detector, small
uncertainties in the detector's parameters 
can have significant effects on the $\gamma$-ray efficiencies.
%The discrepancies have been attributed to the physical characteristics of 
%the detector being different from the manufacturer's specifications.  
%For example, it is common for the physical length of
%the crystal to deviate from the nominal value by a few millimeters.
%For example, the crystal axis may not be parallel to the housing axis,
%For example, the location of the crystal could have an uncertainty of a few 
%millimeters, or the dead layer resulting from a Li-diffused contact could be on the 
%order of a millimeter thicker than specifed. 
%leading to a smaller active region and therefore, smaller efficiencies. 
%Such discrepancies are not unexpected, as the manufacturing process and
%optimization of HPGe detectors are inherently complex.
%The worsening agreement at low energies between simulation and measurement 
%points to an (approximately mm-thick) attenuating material, unspecified by the 
%manufacturer, that was present in the actual detector, but not included in the 
%simulation.
%These errors may seem insignificant, but their effects are noticeable,
%particularly when the source is close to the detector
%and therefore sensitive to small changes in geometry.

%In some experiments (e.g., Ref.~\cite{Boson2008}), better agreement between
%simulations and measurements were attained by x-raying the detector to
%determine the actual (versus nominal) values of the physical parameters. 
%In the present case, x-raying the detector was not an option; therefore,
The adjustments listed in Table~\ref{tab:DetectorParameters} were applied to 
the detector geometry in GEANT4 to make the efficiencies 
from the simulations more closely match those from the benchmarking measurements.  
%lists the detector parameters that were changed.  
The larger disagreement at low energies between the simulated and measured
efficiencies pointed to additional, unspecified attenuating material 
%an approximately mm-thick attenuating material, unspecified by the manufacturer, 
that was present in the actual detector. 
%but not included in the simulation.  
To address this, the thickness of the aluminum mounting cup that immediately 
surrounds the HPGe crystal was increased by 2.25 mm to achieve closer agreement 
between the simulations and measurements.  
%Table~\ref{tab:DetectorParameters} lists each detector parameter that was
%changed.  Column two gives the nominal values of the parameters 
%provided by the manufacturer; column three shows the final values obtained
%after adjusting the detector geometry. 

\begin{figure}[!htbp]
   \includegraphics[width=0.5\textwidth]{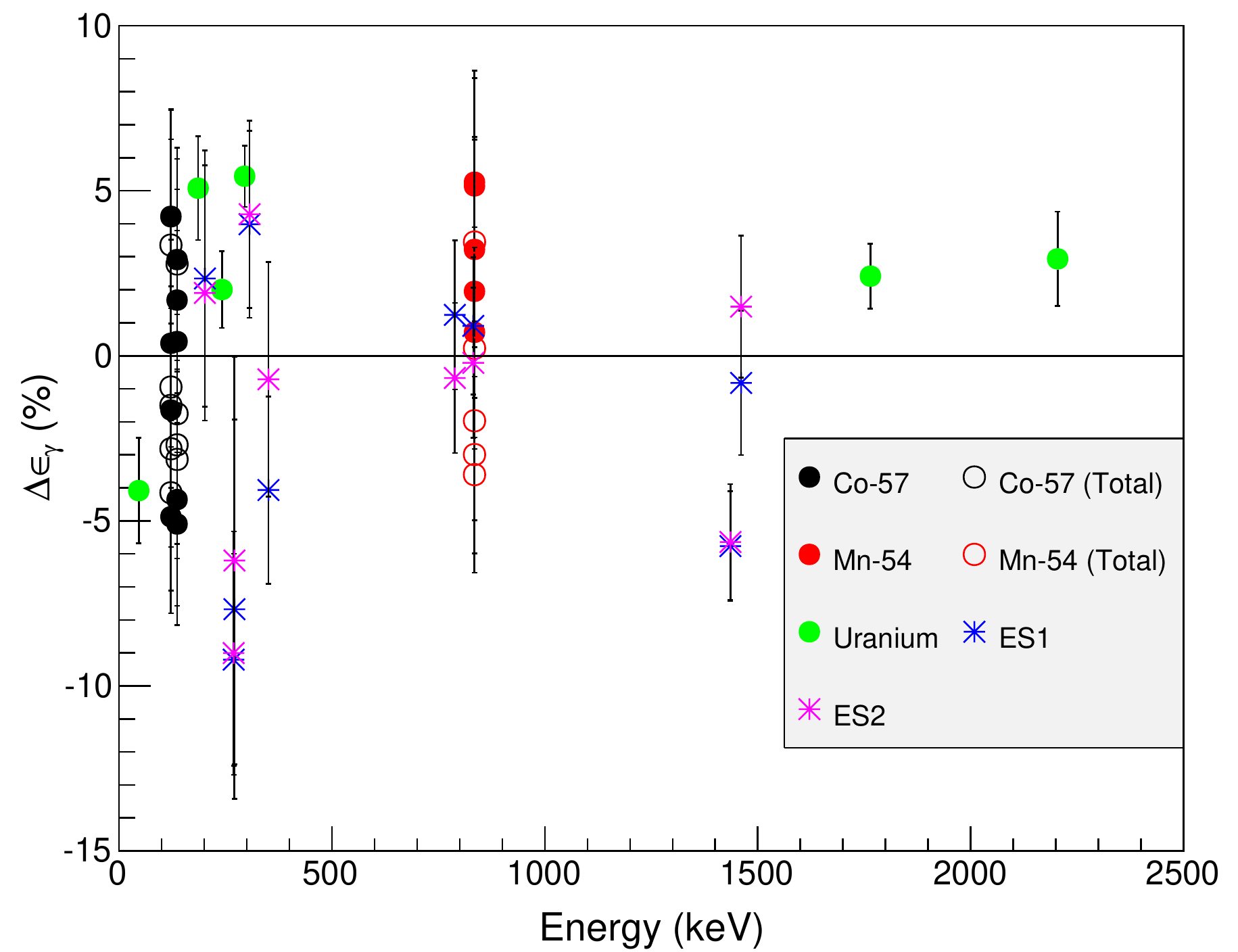}
   \caption
      {
      \label{fig:MerlinSim_FinalDetSpecs}
      Percent differences between the measured and simulated $\gamma$-ray 
      efficiencies as a function of $\gamma$-ray energy.  The simulated 
      efficiencies were obtained using the
      %results for GEANT4 simulations performed 
      adjusted values in Table~\ref{tab:DetectorParameters}.  
      Points corresponding to total efficiencies are indicated with ``(Total)" 
      in the legend.  All other points correspond to peak efficiencies.
      %$\Delta\epsilon_{\gamma}$ is plotted versus $\gamma$-ray 
      %energy for measurements performed with the sources described in
      %Table~\ref{tab:Sources}. 
      %energy for each measurement described in
      %Figure~\ref{fig:SourcePositions}.  The numbers in parentheses 
      %correspond to the point source positions in Figure~\ref{fig:SourcePositions_a}.
      }
\end{figure}

Figure~\ref{fig:MerlinSim_FinalDetSpecs} shows the values of 
$\Delta\epsilon_{\gamma}$ obtained after the adjustments to the detector 
geometry were made in the GEANT4 simulations. 
%versus $\gamma$-ray energy is 
%provided in Figure~\ref{fig:MerlinSim_FinalDetSpecs} for  
%simulations using the set of values in Table~\ref{tab:DetectorParameters},
%column three.
The uncertainties in $\Delta\epsilon_{\gamma}$ take into account the statistical 
uncertainties in the measurements and the simulations, as well as the uncertainties 
in the source activities and branching ratios of the $\gamma$ rays. 
The total uncertainty in the simulated efficiencies was estimated to
be $5 \%$, which is slighly larger than the standard deviation of 
$\Delta\epsilon_{\gamma}$.
%The total uncertainty in the simulated efficiencies was estimated to
%be the standard deviation of $\Delta\epsilon_{\gamma}$, which is $5 \%$.

The photopeak efficiencies of the $\gamma$ rays used to identify the isotopes in
Table~\ref{tab:RadioisotopesObservedInTeO2} were obtained for the irradiated 
TeO\textsubscript{2} powder by performing GEANT4 simulations using the adjusted 
detector values in Table~\ref{tab:DetectorParameters}. 
Simulations indicate that summing could have as much as a 40\% effect for certain 
photopeak efficiencies.  Figure~\ref{fig:MerlinSim_FinalDetSpecs} gives 
confidence that the GEANT4 simulations could model summing correctly and provide
photopeak efficiencies for the irradiated TeO\textsubscript{2} powder with 
around 5\% uncertainty. 

\begin{table*}
\centering
\caption{\label{tab:DetectorParameters} Detector parameters adjusted in 
the GEANT4 simulations.  The nominal values provided by the manufacturer are given, 
along with the values that allowed for satisfactory ($\sim$5\%) agreement 
between the efficiencies from the simulations and the benchmarking measurements.}
\begin{ruledtabular}
\begin{tabular}{l c c}
%   \hline\noalign{\smallskip}
   Parameter & Nominal Value & Adjusted Value \\
    & (mm) & (mm) \\
    %& from Manufacturer & Detector Simulations \\
   \hline\noalign{\smallskip}
   Length of HPGe crystal & 85.5 & 80.5\footnote{The effects of shortening the 
   crystal in the simulation could also be reproduced by using  
   %The effects of shortening the HPGe crystal by 5 mm in the simulation could 
   %also be reproduced by using 
   the nominal length of the crystal and adding a 1.85-mm-thick, 3.5-cm-long 
   copper ring around the aluminum mounting-cup, 8 cm below the top of the
   detector endcap.  Since the presence of such a ring was not specified by the 
   manufacturer and the corresponding simulations provided equivalent results 
   to the shortened crystal geometry, the simulated efficiencies for the  
   80.5-mm-long crystal were used in the cross-section analysis.}\\
   Distance between HPGe crystal and detector window & 0 & 2 \\
   Thickness of aluminum mounting-cup & 0.5 & 2.75 \\
   Thickness of internal dead layer (lithium contact) & 1 & 2 \\
%   \hline
\end{tabular}
\end{ruledtabular}
\end{table*}

\subsection{Neutron transmission during neutron irradiation}
\label{sec:NeutronTransmissionDuringNeutronIrradiation}
Following the neutron irradiation, the aluminum and cadmium foils 
located in front of and behind the TeO\textsubscript{2} powder
were measured with an HPGe detector, 
as was described in Section~\ref{sec:GammaCountingTheIrradiatedTarget}. 
%The spectra for Al1, Al3, Cd1, and Cd2 were used to analyze 
%the neutron transmission through the powder.
%Three of the activation reactions observed in the foils %, along with their
%minimum threshold energies, 
%are listed in Table~\ref{tab:MonitorFoils}.  
%For each activation product, 
%one $\gamma$-ray line in the $\gamma$ spectra was chosen and
%was used to estimate the 
The total neutron transmission, $\bar{T}_{\text{tot}}$, through the 
TeO\textsubscript{2} powder was estimated 
by comparing the activities of the 
activation products in the front foils (Al1 and Cd1) with the activities 
in the back foils (Al3 and Cd2).
%\begin{equation}
%   \label{eq:TotTransmission}
%   \bar{T}_{tot} = \left(\frac{C_{\gamma,b}}{C_{\gamma,f}}\right)\left(\frac{N_{f}}{N_{b}}\right)\left[\frac{\exp(-\lambda t_{s,f}) - \exp(-\lambda t_{e,f})}{\exp(-\lambda t_{s,b}) - \exp(-\lambda t_{e,b})}\right]\left(\frac{\epsilon_{\gamma,f}}{\epsilon_{\gamma,b}}\right).
%\end{equation}
%Here, parameters with subscripts $f$ and $b$ correspond to the front
%and back foils, respectively;
%$\lambda$ is the decay constant for the activation product; and 
%$C_{\gamma,f(b)}$, $N_{f(b)}$, $t_{s,f(b)}$, $t_{e,f(b)}$, and $\epsilon_{\gamma,f(b)}$ 
%are respectively the number of counts in the $\gamma$-ray line, the number of 
%target nuclei in the foil, the start-time of the measurement relative to 
%the end of the irradiation, the end-time of the measurement relative 
%to the end of the irradiation, and the photopeak efficiency of the 
%$\gamma$ ray. 
%Since all of the foils were counted in the same configuration 
%and Al1 (Cd1) has similar dimensions to Al3 (Cd2),  
%$\epsilon_{\gamma,f}$ and $\epsilon_{\gamma,b}$ were assumed to be equal. 
The values of $\bar{T}_{\text{tot}}$ obtained by analyzing the activation 
products are given in Table~\ref{tab:MonitorFoils}, and the uncertainties 
quoted  are statistical.  
%was calculated by adding the uncertainties in $C_{\gamma,f}$, 
%$C_{\gamma,b}$, $N_f$, and $N_b$ in quadrature.  
%Uncertainties in the other parameters in Equation~\ref{eq:TotTransmission} 
%were small enough to be ignored.  
%$\bar{T}$ can then be estimated to be
%
%\begin{equation}
%   \bar{T} \simeq \frac{1+\bar{T}_{tot}}{2}.
%   \label{eq:LANSCEAverageTransmission}
%\end{equation}
%Looking at Table~\ref{tab:MonitorFoils}, one can see that $\bar{T}_{\text{tot}}$ 
%and therefore $\bar{T}$ are 
%is isotope dependent; its value is determined by the 
%product of the neutron-activation cross section and the neutron flux.
To account for the variation in the results, $\bar{T}_{\text{tot}}$ was 
taken to be $0.90 \pm 0.10$, which spans the range of values given in 
Table~\ref{tab:MonitorFoils} along with their uncertainties.  The average 
neutron transmission, $\bar{T}$, through the TeO\textsubscript{2} powder 
was then estimated to be
%estimated using the total transmission through 
%the powder determined in Section~\ref{sec:NeutronTransmissionDuringNeutronIrradiation}. 
%For isotopes with minimum threshold energies greater than 23.4 MeV, 
%$\bar{T}$ was set equal to 1.  For isotopes with minimum threshold energies
%less than 23.4 MeV, $\bar{T}$ was approximated using

\begin{equation}
   \bar{T} \simeq \frac{1 + \bar{T}_{\text{tot}}}{2} = 0.95 \pm 0.05.
\end{equation}

\begin{table}
%\centering
\caption{\label{tab:MonitorFoils} Neutron-transmission results.  Reactions
used to determine the neutron transmission through the TeO\textsubscript{2} powder
are listed, along with the corresponding values of $\bar{T}_{\text{tot}}$.
}
\begin{ruledtabular}
\begin{tabular} {l c }
%   \hline\noalign{\smallskip}
   Reaction \comment{& Minimum Threshold Energy} & $\bar{T}_{\text{tot}}$ \\
    \comment{& (MeV)} &  \\
   \hline\noalign{\smallskip}
   \textsuperscript{27}Al(n,X)\textsuperscript{22}Na \comment{& 23.4} & $0.98 \pm 0.03$ \\
   Cd(n,X)\textsuperscript{105}Ag \comment{& 5.2} & $0.84 \pm 0.01$ \\
   Cd(n,X)\textsuperscript{110m}Ag \comment{& 2.2} & $0.86 \pm 0.01$ \\
%   \hline
\end{tabular}
\end{ruledtabular}
\end{table}

\subsection{Isotope-production rates}
\label{sec:Isotope-productionRates}
The production rate for each isotope can be determined using data
from the $\gamma$-ray spectra collected for the irradiated TeO\textsubscript{2} powder.  
In most cases, the isotope produced in the powder is not
fed by other isotopes during or after the neutron irradiation.  Under this condition, the 
production rate, $R_{\text{30R}}$, can be obtained using:

\begin{equation}
   \label{eq:LANSCEProductionRate_FxnOfGammaCounts}
   R_{\text{30R}} = \frac{\lambda C_{\gamma}}{B_{\gamma}\epsilon_{\gamma}},
   %R_{30R} = \frac{\lambda C_{\gamma}}{B_{\gamma}\epsilon_{\gamma}[\exp(-\lambda t_{s})-\exp(-\lambda t_{e})][1-\exp(-\lambda t_{irrad})]},
\end{equation}
where $\lambda$ is the decay constant of the isotope, 
$C_{\gamma}$ is the number of counts in the 
$\gamma$-ray peak of interest corrected for the growth and decay of the 
isotope during the irradiation and the decay of the isotope after
the irradiation, $B_{\gamma}$ is the branching ratio of the 
$\gamma$ ray, and $\epsilon_{\gamma}$ is the photopeak efficiency of detecting 
the $\gamma$ ray. 
%and $t_{s}$ and $t_{e}$ are 
%respectively the start-time and end-time of the $\gamma$-ray measurement 
%relative to the end of the irradiation.  
The production rates for 
%\textsuperscript{118}Te, 
\textsuperscript{125m}Te and \textsuperscript{127m}Te were described by more 
complex growth-and-decay relations
and were obtained using the appropriate modifications to 
Equation~\ref{eq:LANSCEProductionRate_FxnOfGammaCounts}.

\subsection{Flux-averaged cross sections}
\label{sec:Flux-averagedCross-sections}
The flux-averaged cross section, $\bar{\sigma}_{\text{30R}}$, for neutron
activating an isotope in the irradiated TeO\textsubscript{2} powder
%is defined as
is determined from

\begin{equation}
   %\bar{\sigma}_{\text{30R}} \equiv \frac{\int_{E_{\text{min}}}^{E_{\text{max}}} \sigma(E)\varphi_{\text{30R}}(E)\mathrm{d}E}{\int_{E_{\text{min}}}^{E_{\text{max}}} \varphi_{\text{30R}}(E)\mathrm{d}E},
   \bar{\sigma}_{\text{30R}} = \frac{\int_{E_{\text{min}}}^{E_{\text{max}}} \sigma(E)\varphi_{\text{30R}}(E)\mathrm{d}E}{\int_{E_{\text{min}}}^{E_{\text{max}}} \varphi_{\text{30R}}(E)\mathrm{d}E},
   \label{eq:LANSCEFlux-averagedCross-section}
\end{equation}
where $\sigma(E)$ is the cross section for producing the isotope
with neutrons of kinetic energy $E$, %interact with target nuclei in the 
%TeO\textsubscript{2} powder, 
$\varphi_{\text{30R}}(E)$ is the %30R 
differential neutron flux hitting the front of the target in 
units of [neutrons/(cm\textsuperscript{2}$\cdot$s$\cdot$MeV)],
and $E_{\text{min}}$ and $E_{\text{max}}$ are respectively the lowest and highest neutron
energies hitting the TeO\textsubscript{2} powder. %was exposed to.  

The isotope-production rate %from Section~\ref{sec:Isotope-productionRates}
can also be expressed as

%\begin{eqnarray}
\begin{equation}
   \label{eq:LANSCEProductionRate_2}
   R_{\text{30R}} \approx N \bar{\sigma}_{\text{30R}} \bar{T} \int_{E_{\text{min}}}^{E_{\text{max}}} \varphi_{\text{30R}}(E) \mathrm d E, 
\end{equation} 
%\end{eqnarray} 
where %$\bar{T}$ represents the average neutron transmission through the 
%powder and 
$N$ is the number of tellurium nuclei in the powder  
(except for the production of \textsuperscript{7}Be, where
$N$ is the number of oxygen nuclei in the powder). 
%because 
%\textsuperscript{7}Be is produced almost exclusively by neutron 
%interactions with oxygen.  
%For all other radioisotopes activated in the powder, $N$ is the number of tellurium nuclei.  

%For the 30R neutron beam, neutron kinetic energies ranged from thermal up 
%to 800 MeV;
%%Setting $E_{min}$ to any energy less than 1.25 MeV is an issue, however,
%%because 
%however, the neutron flux during the experiment was not recorded for 
%kinetic energies below 1.25 MeV.  
%It is also unfavorable to evaluate cross sections for isotopes that were
%produced by a significant number of low-energy-neutron interactions 
%($\ltaeq 0.1$ MeV kinetic energy), as these neutrons were primarily 
%generated by higher energy neutrons slowing down in nearby materials. 
The total neutron flux below 1.25 MeV during the irradiation was determined to be 
nonnegligible from 
%by using the $\gamma$-ray measurements of the gold foils to
%analyze 
the amount of \textsuperscript{198}Au created 
by (n,$\gamma$) reactions in the gold foils. 
Therefore, cross sections could be obtained only for isotopes produced soley (or primarily) by
interactions with neutrons of energy $>1.25$ MeV, and 
%for isotopes with low probabilities
%of being produced from interactions with $< 1.25$-MeV neutrons, and 
in these cases, $E_{\text{min}}$ and $E_{\text{max}}$ from Equation~\ref{eq:LANSCEProductionRate_2}
could be set to 1.25 MeV and 800 MeV, respectively.
%and cross-sections were not obtained for isotopes with high 
%probabilites of being produced from interactions with $< 1.25$-MeV neutrons. 
%Section~\ref{sec:NeutronFluxBelow1.25MeV} discusses how the total flux
%below 1.25 MeV was estimated using the two gold foils in the TeO\textsubscript{2}
%target.  The flux was determined to be non-negligible, and therefore 

%Fortunately, for most of the isotopes, %produced in the TeO\textsubscript{2} powder, 
%the neutron-interaction cross section below 1.25 MeV is either 
%equal to zero or negligible.  In these cases, we set $E_{min} = 1.25$ MeV, 
%$E_{max} = 800$ MeV, and rewrite Equation~\ref{eq:LANSCEProductionRate_2} as  
%\ref{eq:LANSCEFlux-averagedCross-section} 

%\begin{eqnarray}
%%   \bar{\sigma}_{30R} &\approx& \frac{\int_{1.25\text{ MeV}}^{800\text{ MeV}} \sigma(E)\varphi_{30R}(E)\mathrm{d}E}{\int_{1.25\text{ MeV}}^{800\text{ MeV}} \varphi_{30R}(E)\mathrm{d}E},\\
%  R_{30R} &\approx& N \bar{\sigma}_{30R} \bar{T} \int_{1.25\text{ MeV}}^{800\text{ MeV}} \varphi_{30R}(E) \mathrm{d}E. \label{eq:LANSCEProductionRate_Final}
%\end{eqnarray}
The flux-averaged cross sections, shown in Table~\ref{tab:RadioisotopesObservedInTeO2},
can then be determined from Equations~\ref{eq:LANSCEProductionRate_FxnOfGammaCounts}
and~\ref{eq:LANSCEProductionRate_2}.

\section{Comparing measured and calculated cross sections}
%The isotope-production cross sections were also calculated with the ACTIVIA code 
%by implementing the S\&T semi-empirical formulae. 
%Isotope-production cross sections calculated using the S\&T semi-empirical formulae
%were also obtained with the ACTIVIA code.
%Calculations based on the S\&T semi-empirical formulae were also performed with 
%the ACTIVIA code to obtain isotope-production cross sections for tellurium.
Isotope-production cross sections for tellurium were also obtained by using the
ACTIVIA code to perform calculations based on the S\&T semi-empirical formulae.
These formulae were originally developed to describe proton-nucleus interactions,
but they are assumed to be applicable to neutron-nucleus interactions
as well.
The calculated cross sections are reported in Table~\ref{tab:RadioisotopesObservedInTeO2}.  
Although the formulae are only valid for proton and neutron energies $\geq 100$ MeV 
and they do not distinguish between ground and metastable states in product nuclei,
the calculated and measured cross sections agree reasonably well, within a factor of
3 on average.  One should note that the cross section calculated for 
\textsuperscript{110m}Ag was underestimated by approximately a factor of 5.

\section{Cosmogenic-activation background in the CUORE experiment}
The CUORE experiment will use an array of 988 high-resolution, 
low-background TeO\textsubscript{2} bolometers to search for the 
$0\nu\beta\beta$ decay of \textsuperscript{130}Te.
Each bolometer is comprised of a 5$\times$5$\times$5 cm\textsuperscript{3} natural-TeO\textsubscript{2} 
crystal that serves as both a source and a detector of the decay. 
CUORE is aiming for a background rate of $10^{-2}$ counts/(keV$\cdot$kg$\cdot$ y)
at the \textsuperscript{130}Te Q\textsubscript{$\beta\beta$} value of 2528 keV,
which would allow the experiment to reach a 
%1$\sigma$ half-life sensitivity of $1.6\times10^{26}$ y after several years of running 
half-life sensitivity of $9.5\times10^{25}$ years ($90 \%$ C.L.), assuming a 
live time of 5 years and a full-width-at-half-maximum energy resolution of 5 keV 
\cite{Alessandria2011}.

Using the results of the neutron-activation measurement discussed in this work 
and the proton-activation measurements of Ref.~\cite{Barghouty2013}, one
can determine the background contribution to CUORE from the cosmogenic 
activation of the TeO\textsubscript{2} crystals that occurs during sea 
transportation from the crystal-production site in Shanghai, China to LNGS 
in Italy.
%Cosmogenic activation can be problematic because the radionuclides produced
%constitute an intrinsic contamination of the crystals.  
%Radioactivity present inside the crystals cannot be removed or shielded 
%during counting.
%According to Table~\ref{tab:RadioisotopesObservedInTeO2} and 
%Ref.~\cite{Barghouty2013}, \textsuperscript{110m}Ag and \textsuperscript{60}Co 
%are the only two long-lived radioisotopes able to contribute background at the 
%$0\nu\beta\beta$-decay peak due to their Q values being greater than the 
%%\textsuperscript{130}Te 
%Q\textsubscript{$\beta\beta$} value of 2528 keV.
The results of both this work and Ref.~\cite{Barghouty2013} indicate that 
\textsuperscript{110m}Ag and \textsuperscript{60}Co 
are the only two long-lived radioisotopes that will contribute meaningfully to the
background at the $0\nu\beta\beta$-decay peak due to their Q values being greater than 
Q\textsubscript{$\beta\beta$}.
\textsuperscript{110}Ag will also contribute a small amount to the background 
because \textsuperscript{110m}Ag decays to it 1.33\% of the time.  
%it is in secular equilibrium with its parent \textsuperscript{110m}Ag. 
%in bold font all 
%have Q-values (column six) greater than 2528 keV, which means they can 
%contribute events to the $0\nu\beta\beta$ decay region.  Looking at the 
%half-lives, however, one can see that most of these isotopes have short 
%half-lives and will not be present once CUORE begins counting.  

The production rates, $R$, of \textsuperscript{110m}Ag and 
\textsuperscript{60}Co were each estimated to be

\begin{equation}
  R \approx N\sum_{i}\sigma_{i}\phi_{\text{CR},i},
  \label{eq:ProductionRateInCUORE}
\end{equation}
where $\sigma_{i}$ is the isotope-production cross section assigned to 
energy bin $i$, and $\phi_{\text{CR},i}$ is the differential cosmic-ray 
neutron flux at sea-level integrated over energy bin $i$.  
The energy bins, integrated fluxes, and $\sigma_{i}$
values are given in Table~\ref{tab:CUOREProductionRateParameters}. 
The cosmic-ray neutron flux determined by Gordon \textit{et al.} 
\cite{Gordon2004} was used in this analysis, with the 
parameter $F_{\text{BSYD}}$ from Ref.~\cite{Gordon2004} taken
to be $0.73 \pm 0.22$ \cite{Wang2014} for the route used to ship the 
TeO\textsubscript{2} crystals.  
%According to Table~\ref{tab:CUOREProductionRateParameters},
80\% of the \textsuperscript{110m}Ag and as much as 37\% of the 
\textsuperscript{60}Co were produced by 1.25--800 MeV neutrons.    

The fraction of \textsuperscript{110m}Ag, \textsuperscript{110}Ag,
and \textsuperscript{60}Co decays that deposit energy in a 60 keV-wide
region-of-interest (ROI) surrounding the $0\nu\beta\beta$-decay peak was estimated 
using GEANT4 simulations of a single 5$\times$5$\times$5 cm\textsuperscript{3}
TeO\textsubscript{2} crystal.  
The values obtained were $0.5 \%$, $0.4 \%$, and $1 \%$ for \textsuperscript{110m}Ag,
\textsuperscript{110}Ag, and \textsuperscript{60}Co decays, respectively.
%The efficiencies for \textsuperscript{110m}Ag, \textsuperscript{110}Ag, and
%\textsuperscript{60}Co to deposit energy in the ROI
In the full CUORE array, the presence of nearby crystals would often lead 
to energy being deposited in more than one crystal.  As most 
$0\nu\beta\beta$-decays would deposit all of their energy in a single
crystal, the background can be reduced by rejecting events in which
energy was deposited in more than one crystal.
%These simulations neglected the CUORE array's
%ability to veto coincidence events between multiple crystals. However, 
%vetoing is expected to have a small impact on the efficiencies because the 
Simulations of a $3\times3\times3$ TeO\textsubscript{2}-crystal array 
%performed after adding nearest-neighbor crystals to the single-crystal
indicate that rejecting multi-crystal events can suppress the \textsuperscript{110m}Ag
contribution to the ROI by a factor of $\sim$2, while the contributions
from \textsuperscript{110}Ag and \textsuperscript{60}Co will be minimally affected. 
%However, the decay Q values for \textsuperscript{110m}Ag, 
%\textsuperscript{110}Ag, and \textsuperscript{60}Co are only a few hundred 
%keV higher than Q\textsubscript{$\beta\beta$}. 
%Therefore, the rejection of multi-crystal events is expected to have only
%a small impact on the background contribution at 
%Q\textsubscript{$\beta\beta$} from these isotopes, as there is little
%ionizing-radiation energy available to be deposited in a second crystal. 
%leaving very few $\gamma$ rays to interact with other crystals after 
%$\sim$2528 keV is deposited in the first crystal. 

%Then, assuming each crystal spends $\sim$3 months at sea level and $\sim$5
%years underground before the CUORE experiment begins operation in late 2015, 
To estimate the background rate at Q\textsubscript{$\beta\beta$} from
cosmogenic activation of TeO\textsubscript{2}, the
following assumptions were made: (1) each crystal spends 3 months at sea level, 
(2) no \textsuperscript{110m}Ag, \textsuperscript{110}Ag, and 
\textsuperscript{60}Co were present at the beginning of shipment due to their 
removal during the crystal-growth process, and (3) crystals were delivered to 
LNGS and stored underground at a constant rate from early 2009 to late 2013 \cite{Wang2014}.
%before the CUORE experiment begins operation in late 2015, 
%on January 1, 2016,
%and CUORE begins operation on January 1, 2016 -- January 1, 2021, 
The resulting contamination levels for
\textsuperscript{110m}Ag+\textsuperscript{110}Ag and \textsuperscript{60}Co
when CUORE begins operation in late 2015 will be 
$\sim$$2\times10^{-8}$ Bq/kg and $\sim$$10^{-9}$ Bq/kg, respectively, which 
correspond to background rates of $\sim$$6\times 10^{-5}$ counts/(keV$\cdot$kg$\cdot$ y)
and $\sim$$7\times10^{-6}$ counts/(keV$\cdot$kg$\cdot$ y), respectively.
%the \textsuperscript{110m}Ag+\textsuperscript{110}Ag and \textsuperscript{60}Co
%contamination levels when CUORE begins operation in late 2015 will be 
%$\sim$$2\times10^{-8}$ Bq/kg and $\sim$$10^{-9}$ Bq/kg, respectively, which 
%correspond to background rates of $\sim$$6\times 10^{-5}$ counts/(keV$\cdot$kg$\cdot$ y)
%and $\sim$$7\times10^{-6}$ counts/(keV$\cdot$kg$\cdot$ y), respectively.
%the \textsuperscript{110m}Ag+\textsuperscript{110}Ag and \textsuperscript{60}Co
%background rates in counts/(keV$\cdot$kg$\cdot$ y) in the 60-keV-wide 
%$0\nu\beta\beta$-decay ROI were determined to be 
%$\sim$$3\times 10^{-5}$ and $\sim$$6\times10^{-6}$, respectively, 
%$\sim$$6\times 10^{-5}$ ($\sim$$2\times10^{-8}$ Bq/kg \textsuperscript{110m}Ag+\textsuperscript{110}Ag 
%contamination level) and $\sim$$7\times10^{-6}$ ($\sim$$10^{-9}$ Bq/kg
%\textsuperscript{60}Co contamination level), respectively, 
%at the beginning of counting in late 2015.  
%After 5 years of running, the rates will decrease to 
%$\sim$$4 \times 10^{-7}$ ($\sim$$10^{-10}$ Bq/kg) and  $\sim$$4 \times 10^{-6}$
%($\sim$$6\times10^{-10}$ Bq/kg), respectively.
After 5 years of running, the contamination levels will decrease to 
$\sim$$2\times10^{-10}$ Bq/kg for \textsuperscript{110m}Ag+\textsuperscript{110}Ag and 
$\sim$$6\times10^{-10}$ Bq/kg for \textsuperscript{60}Co, which correspond to
background rates of $\sim$$4 \times 10^{-7}$ counts/(keV$\cdot$kg$\cdot$ y) and 
$\sim$$4 \times 10^{-6}$ counts/(keV$\cdot$kg$\cdot$ y), respectively.
The contamination levels given here are lower than those predicted in 
Ref.~\cite{Lozza2014} due to Lozza \textit{et al.} assuming a longer exposure time of 1 year 
and a shorter overall cooling time underground of 2 years.
%, at the end of counting.  
Rejecting multi-site events should decrease the 
\textsuperscript{110m}Ag+\textsuperscript{110}Ag background rates by a factor 
of $\sim$2.
Although the background rates in the ROI are at least two orders of magnitude lower than
the current CUORE goal background of $10^{-2}$ counts/(keV$\cdot$kg$\cdot$ y), 
%cosmogenic activation of TeO\textsubscript{2}
%will not be an issue.  However, 
for future experiments striving for essentially zero background, cosmogenic activation 
may have to be addressed more stringently.

\begin{table*}
%\centering
\caption{\label{tab:CUOREProductionRateParameters} Energy bins used in
the estimation of the \textsuperscript{110m}Ag and \textsuperscript{60}Co
production rates.  The differential cosmic-ray neutron flux at sea-level
integrated over each bin is provided.  The isotope-production cross 
sections assigned to each bin are also listed.  For bin 1, the cross 
sections obtained in this work are used.  For bins 2, 3, and 4, the cross 
sections used were those measured in proton-activation experiments  
with 800 MeV, 1.4 GeV, and 23 GeV protons respectively.  The individual
contributions to $R$ in units of [s\textsuperscript{-1}] and [\%] are 
given in the last two columns.
}
\begin{ruledtabular}
\begin{tabular} {c c c c c c c}
%   \hline\noalign{\smallskip}
   Bin & Bin Range & Integrated Neutron Flux & \multicolumn{2}{c}{Cross section} & \multicolumn{2}{c}{Contribution to $R$} \\
    &  & (s\textsuperscript{-1}$\cdot$cm\textsuperscript{2}) & \multicolumn{2}{c}{(mb)} & \multicolumn{2}{c}{(s\textsuperscript{-1})} \\
   \hline\noalign{\smallskip}
      & & & \textsuperscript{110m}Ag & \textsuperscript{60}Co & \textsuperscript{110m}Ag & \textsuperscript{60}Co\\
   \hline\noalign{\smallskip}
    1 & 1.25 MeV -- 800 MeV & $(3.7 \pm 1.3)\times10^{-3}$ & $0.28 \pm 0.04$ & $< 0.0016$ & $(2.9 \pm 1.1)\times10^{-6}$ & $<(1.7 \pm 0.6)\times10^{-8}$  \\
     & & & & & (80\%) & ($<37$\%)  \\
     & & & & & & \\
    2 & 800 MeV -- 1.4 GeV & $(5.3 \pm 1.9)\times10^{-5}$ & $3.95 \pm 0.40$\footnote{\label{800MeVXSec}The value of this cross section was reported incorrectly
in Ref.~\cite{Barghouty2013}, but correctly in Ref.~\cite{Quiter2005}.} \cite{Quiter2005} & $0.09 \pm 0.04$ \cite{Barghouty2013} & $(5.9 \pm 2.2)\times10^{-7}$ & $(1.4 \pm 0.8)\times10^{-8}$ \\
     &  &  &  &  & (16\%) & ($>30$\%) \\
     & & & & & & \\
    3 & 1.4 GeV -- 23 GeV & $(2.6 \pm 1.0)\times10^{-5}$ & $1.9 \pm 0.3$ \cite{Barghouty2013} & $0.20 \pm 0.04$ \cite{Barghouty2013} & $(1.4 \pm 0.6)\times10^{-7}$ & $(1.5 \pm 0.6)\times10^{-8}$ \\
    &  &  &  &  & (3.9\%) & ($>33$\%) \\
     & & & & & & \\
    4 & 23 GeV -- 150 GeV & $(1.6 \pm 0.6)\times10^{-7}$ & $0.88 \pm 0.59$ \cite{Barghouty2013} & $0.75 \pm 0.08$ \cite{Barghouty2013} & $(4.0 \pm 3.1)\times10^{-10}$ & $(3.4 \pm 1.3)\times10^{-10}$ \\
    & &  &  &  &  (0.01\%) &  ($>0.8$\%) \\
%   \hline\noalign{\smallskip}
%   \textsuperscript{60}Co & 1 & 1.25 MeV -- 800 MeV & $0.0012 \pm 0.0001$ \\
%    & 2 & 800 MeV -- 1.4 GeV & $0.09 \pm 0.04$ \cite{Barghouty2013}\\
%    & 3 & 1.4 GeV -- 23 GeV & $0.20 \pm 0.04$ \cite{Barghouty2013} \\
%    & 4 & 23 GeV -- 150 GeV & $0.75 \pm 0.08$ \cite{Barghouty2013}\\
%   \hline
\end{tabular}
\end{ruledtabular}
\end{table*}

\section{Conclusions}
Flux-averaged cross-sections for cosmogenic-neutron activation of 
radioisotopes in natural tellurium were measured by irradiating 
TeO\textsubscript{2} powder with a neutron beam containing neutrons of 
kinetic energies up to $\sim$800 MeV, and having an energy spectrum 
similar to that of cosmic-ray neutrons at sea-level.
%These cross-sections were compared with results from semi-empirical 
%cross-section calculations using the ACTIVIA code.
The cross sections obtained for \textsuperscript{110m}Ag and 
\textsuperscript{60}Co, the two isotopes which have both half-lives of
order a year or longer and Q values larger than the 
Q\textsubscript{$\beta\beta$} of \textsuperscript{130}Te, were combined 
with results from tellurium 
activation measurements with 800 MeV -- 23 GeV protons to estimate the
background in the CUORE experiment from cosmogenic activation of the
TeO\textsubscript{2} crystals. 
The anticipated \textsuperscript{110m}Ag+\textsuperscript{110}Ag and \textsuperscript{60}Co
background rates in [counts/(keV$\cdot$kg$\cdot$ y)] at the 
$0\nu\beta\beta$-decay peak were determined to be $\sim$$6\times 10^{-5}$ and
$\sim$$7\times10^{-6}$, respectively, at the beginning of counting and
$\sim$$4 \times 10^{-7}$ and  $\sim$$4 \times 10^{-6}$, respectively, after 5
years of counting. The \textsuperscript{110m}Ag+\textsuperscript{110}Ag rates
should decrease by a factor of $\sim$2 if multi-crystal events are efficiently rejected. 
These rates are at least two orders of magnitude lower than the 
goal background for the CUORE experiment.
%As these contributions are at least two orders of magnitude lower than
%CUORE's goal background, cosmogenic activation of TeO\textsubscript{2}
%will not be an issue.  However, for future experiments striving for 
%zero background, cosmogenic activation will have to be addressed more
%stringently.  

\section{Acknowledgements}
We gratefully acknowledge the many valuable discussions with
Maura Pavan and Silvia Capelli from the CUORE Collaboration.  
This work was supported by Lawrence Livermore National Laboratory under 
Contract DE-AC52-07NA27344, Los Alamos National Laboratory under Contract 
DE-AC52-06NA25396, Lawrence Berkeley National Laboratory under Contract 
DE-AC02-05CH11231, the U.S. Department of Energy Office of Defense Nuclear 
Nonproliferation (NA-22), the U.S. Department of Energy National Nuclear 
Security Administration under Award Number DE-NA0000979,
and the Nuclear Forensics Graduate Fellowship from the U.S. 
Department of Homeland Security under Grant Award Number 2012-DN-130-NF0001-02. 
The views and conclusions contained in this document are those of the 
authors and should not be interpreted as necessarily representing the 
official policies, either expressed or implied, of the U.S. Department of 
Homeland Security.
\bibliography{Bibliography}

\end{document}